\documentclass{aa}

\usepackage{graphicx}  
\usepackage{txfonts}

\usepackage{float}
\usepackage{multicol}
\usepackage{caption}
\usepackage{subcaption}
\usepackage{geometry}
\usepackage{setspace}

\usepackage[colorlinks=true,linkcolor=webgreen,citecolor=webblue,bookmarks=true]{hyperref}

\usepackage{latexsym}
\usepackage{wrapfig}

\usepackage[usenames]{color}
\definecolor{webblue}{rgb}{0,0,.5}
\definecolor{webgreen}{rgb}{0,.5,0}
\usepackage{chngcntr}
\usepackage{ulem}

\begin{document}
        
        \title{Characterising atmospheric gravity waves on the nightside lower clouds of Venus: A systematic analysis}

        \author{J.E. Silva\inst{1}
        \and P. Machado\inst{1}
        \and J. Peralta\inst{2} 
        \and F. Brasil\inst{1}
        \and S. Lebonnois\inst{3}
        \and M. Lefèvre\inst{4}
                        }
        \institute{Institute of Astrophysics and Space Sciences, Observat\'orio Astron\'omico de Lisboa, Ed. Leste, Tapada da Ajuda, 1349-018 Lisbon, Portugal
        \and Institute of Space and Astronautical Science, Japan Aerospace Exploration Agency - 3-1-1, Yoshinodai, Chuo-ku, Sagamihara, Kanagawa 252-5210, Japan
        \and Laboratoire de Météorologie Dynamique (LMD/IPSL), Sorbonne Université, Centre National de la Recherche Scientifique, École Polytechnique Normale Supérieure, Paris, France
        \and Department of Physics (Atmospheric, Oceanic and Planetary Physics), University of Oxford, Parks Rd, Oxford, OX1 3PU, UK
        }

        \date{Received December 21, 2020}
        

        



\abstract{
        We present the detection and characterisation of mesoscale waves on the lower clouds of Venus using images from the Visible Infrared Thermal Imaging Spectrometer (VIRTIS-M) onboard the European Venus Express space mission and from the 2 $\mu$m camera (IR2) instrument onboard the Japanese space mission Akatsuki. We used image navigation and processing techniques based on contrast enhancement and geometrical projections to characterise morphological properties of the detected waves, such as horizontal wavelength and the relative optical thickness drop between crests and troughs. Additionally, we performed phase velocity and trajectory tracking of wave packets. We combined these observations to derive other properties of the waves such as the vertical wavelength of detected packets. Our observations include 13 months of data from August 2007 to October 2008, and the entire available data set of IR2 from January to November 2016. \\
        We characterised almost 300 wave packets across more than 5500 images over a broad region of the globe of Venus. Our results show a wide range of properties and are not only consistent with previous observations but also expand upon them, taking advantage of two instruments that target the same cloud layer of Venus across multiple periods. In general, waves observed on the nightside lower cloud are of a larger scale than the gravity waves reported in the upper cloud.\\
        This paper is intended to provide a more in-depth view of atmospheric gravity waves on the lower cloud and enable follow-up works on their influence in the general circulation of Venus.
        }

        
        \titlerunning{Characterising Atmospheric Waves on the Nightside of Venus}
        \maketitle

        \section{Introduction}
        
                An atmospheric internal gravity wave is an oscillatory disturbance on an atmospheric layer in which the buoyancy of the displaced air parcels acts as the restoring force. As such, this kind of wave can only exist in a continuously stably stratified atmosphere, that is, a fluid in which the static stability is positive and horizontal variations in pressure (within the atmospheric layer) are negligible when compared to the vertical variations (in altitude) \citep{Sutherland2010}.\\
                These waves represent an efficient transport mechanism of energy and momentum which can dissipate at different altitudes and force the dynamics of several layers of the atmosphere. This dissipation or wave breaking can dump the transported momentum and energy to the mean flow, contributing to an acceleration, thus significantly altering the thermal and dynamical regime of the atmosphere \citep{Alexander2010}.\\
                The attributes of internal gravity waves are of particular importance in the case of Venus, a planet with a zonal retrograde super-rotating atmosphere with winds up to 60 times faster than the solid globe of the planet \citep{SanchezLavegabook}. The mechanism that drives this motion has been the subject of debate for over 40 years. Multiple modelling attempts and observations of cloud dynamics  both from the ground and space have been made, and still there is no complete answer \citep{SanchezLavega2017}.\\
                According to the most recent models and observations, super-rotation starts to become more prominent at cloud level ($\sim$ 45 km), with steadily increasing wind speeds up to the top of the clouds of Venus ($\sim$ 67-70 km), where they can reach 100 m/s \citep{SanchezLavega2008}. The proper mechanism that powers the zonal flow to such magnitudes is still unclear but one possible contribution could be in the form of atmospheric gravity waves generated from a convective layer in the middle-bottom of the clouds that can transport momentum upwards, powering such winds \citep{HouFarrel1987, Peralta2008}. Furthermore, measurements from several \textit{in situ} probes as well as recent radio-occultation data analysis from the Japanese Aerospace Exploration Agency (JAXA) Akatsuki space mission show that Venus features a convective zone in the middle of the cloud layer (at a height of $\sim$ 49 -55 km) where heat, momentum, and chemical species are mixed and change the dynamical regime of the atmosphere considerably \citep{Zasova2007, Tellmann2009, Ando2020}.\\
                Venus' atmosphere displays an incredible variety of waves, which are detected at different wavelengths ranging from the ultraviolet to the near-infrared. The periodic structure of the waves observed in the ultraviolet can be seen as differences in the reflected light at the top of the clouds while waves in the infrared appear through opacity patterns to the thermal radiation under the cloud layer where they propagate \citep{Belton1976, Rossow1990, Peralta2008, Piccialli2014}. Observations of waves in the upper cloud with the Venus Monitoring Camera (VMC) instrument onboard the European Space Agency (ESA) Venus Express (VEx) space mission lead to the detection of periodic structures interpreted as gravity waves whose activity was mostly limited to the cold collar region (60$^{\circ}$ - 80$^{\circ}$) and concentrated above a mountainous region on the northern hemisphere (Ishtar Terra)\citep{Piccialli2014}. This suggested these waves were generated by a Kelvin-Helmholtz instability or that waves are excited by the interaction of the lower atmosphere with the surface topography. However, a need for improving the statistics of wave analysis to further develop these and other hypotheses has been expressed \citep{Piccialli2014}. Additional observations of the upper clouds using Venus Infrared Thermal Imaging Spectrometer - Mapper (VIRTIS-M) images at target wavelengths 3.9 $\mu$m and 5 $\mu$m reveal the presence of a large number of stationary mesoscale waves \citep{Peralta2017b}. Interestingly, as these wavelengths target the nightside upper clouds and the waves seem related to topography, the same structures are mysteriously missing on the nightside lower clouds. On the other hand, waves previously detected in the lower cloud of Venus show a large diversity of properties and morphologies which span an extended latitudinal range (40$^\circ$ - 75$^{\circ}$) limited to the southern hemisphere. The location of these waves seems uncorrelated to any notable topographical feature on the surface of Venus or the local time \citep{Peralta2008}.\\
                Radio occultation data also allow direct detection of waves through small-scale temperature fluctuations, the latest studies of which reveal significant wave activity in the cold collar region favouring the northern hemisphere \citep{Tellmann2012}. 
                With these results, \citet{Tellmann2012} concludes that waves are generated by either convection or topographical forcing, supporting other wave studies using different observation techniques \citep{Peralta2008, Piccialli2014}.\\
                Here, we present the results of a systematic search for wave-like features on the lower clouds of Venus' nightside (estimated to be in a region between $\sim$ 44 and 49 km above the surface) using imaging data from two different instruments, namely the VIRTIS instrument \citep{Drossart2007} onboard Venus Express \citep{Svedhem2007} and the 2-$\mu$m camera (IR2) onboard Akatsuki \citep{Nakamura2016}.
        
        \section{Data acquisition}
        \label{DataAcquisition}
        
        VIRTIS is an instrument equipped with two separate telescopes which work on two channels: VIRTIS-M a mapping spectrometer that operates in two wavelength ranges (VIRTIS-M-VIS from 0.3 to 1 $\mu$m and VIRTIS-M-IR from 1 to 5 $\mu$m); and VIRTIS-H, a high-resolution spectrometer focused on the infrared \citep{Drossart2007}. In imaging mode, VIRTIS extracts cubes of images, dividing its spectral range into 432 wavelength slits. This setup allows simultaneous visualisation of different layers of the atmosphere. \citet{SanchezLavega2008} provide a relation between wavelength and cloud altitude determination as well as a detailed account of the cloud dynamics of Venus. \citet{Peralta2017} also give an overview of the spectral regions that target specific cloud layers of the atmosphere.\\
        Venus Express featured a highly elliptical orbit throughout its mission, with an apocenter at 60,000 km from the planet and a pericenter as close as 350 km above the cloud tops facing approximately 80$^{\circ}$N of the planet. Because of the long integration times of the VIRTIS instrument, the spacecraft would move too fast as it approached the pericenter for effective mapping of the disc to construct an image cube, and therefore most VIRTIS observations are focused on the southern hemisphere of Venus.\\
        We set out to follow up on the analysis performed in \citet{Peralta2008} and conducted our search for waves over 13 months of VIRTIS data starting in August 2007 until the unfortunate malfunction of the infrared (IR) channel of VIRTIS-M in October 2008, which disabled its use for the rest of the mission \citep{Hueso2012}. However, we do include the data from \citet{Peralta2008} in our results, who observed the lower cloud from April 2006 to March 2007, in order to present the data set in its most complete format. A total of 239 orbits were examined, each image at four target wavelength ranges (1.74, 2.25, 3.9 and 5 $\mu$m). The first two wavelengths are sensitive to the opacity of the lower cloud layer at a height of $\sim$ 44-48 km against the brighter background thermal radiation from below, while the last two sense the thermal emission of the upper clouds \citep{Peralta2017}. As this study focuses on the lower cloud of Venus, observations in the latter wavelengths are left to a future publication.\\
        The Akatsuki space mission was launched in March 2010 and reached the vicinity of Venus in December of the same year, but the orbit insertion manoeuvre failed. However, 5 years later a new opportunity came for orbit reinsertion which was fortunately successful \citep{Nakamura2016}. This led to a more elongated orbit than previously planned, but it allowed for more continuous observations because the apparent disc of Venus is found within the field of view (FOV) about 96\% of the time in one orbital revolution compared to the originally planned 60\% \citep{Nakamura2016}. The IR2 camera operates at three wavelengths for standard observation: 1.73, 2.26, and 2.32 $\mu$m. The first two are affected only by $CO_2$ absorption while the last one also features an absorption band from CO. Using these wavelengths, the outgoing radiation originates from altitudes of between 35 and 50 km \citep{Nakamura2011, Satoh2016}.\\
        The orbital characteristics after orbit reinsertion, and particularly observations of the full apparent disc of Venus for longer time intervals than was possible for VIRTIS observations, opened up equatorial latitudes for exploration of atmospheric waves, enabling a more complete characterisation of different locations in the atmosphere of Venus. Images at 2.26 $\mu$m were predominantly used for the IR2 data because they offered the best conditions to precisely detect and characterise features, and more importantly minimised light pollution from the dayside of the disc of Venus caused by multiple reflections of infrared light on the detector \citep{Satoh2017}. We inspected all IR2 data along 30 orbits, from March 2016 to December 2016, before the acquisition of images was indefinitely interrupted \citep{Iwagami2016}. In total, 1255 VIRTIS-M images, each at the target wavelengths mentioned above, and 1639 IR2 images were analysed to detect atmospheric waves during the periods selected.
        
        \section{Methods}
                \subsection{Detection}
\begin{figure*}[!]
                \centering
                \includegraphics[width=0.8\paperwidth]{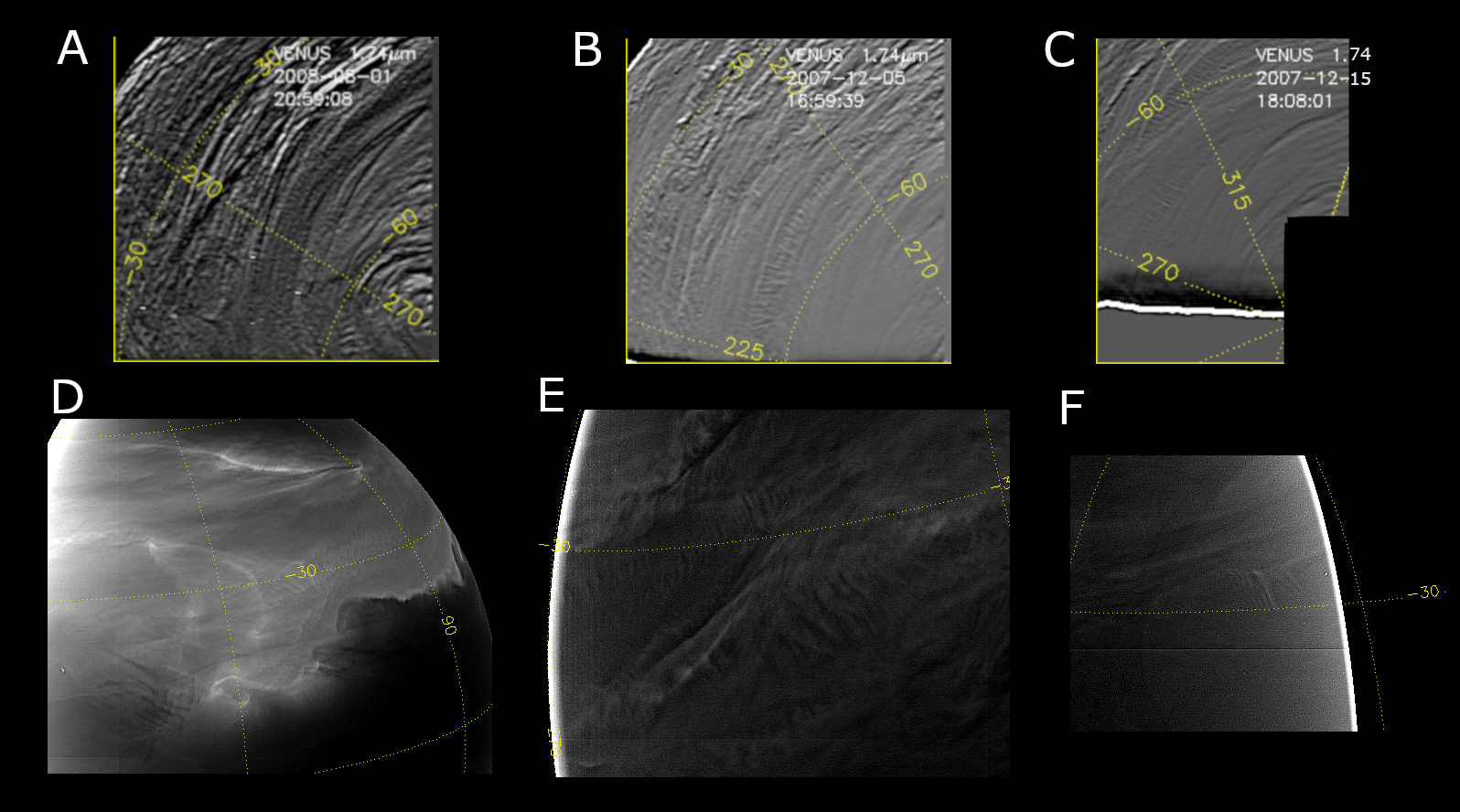}
                \caption{\footnotesize Examples of detected waves on navigated images. A-C: VIRTIS images processed with a directional kernel and unsharp masking; D-F: IR2 images processed with unsharp masking and histogram equalisation.}
                \label{wave_examples}
        \end{figure*}

\begin{figure}[!]
                \centering\includegraphics[width=0.8\columnwidth]{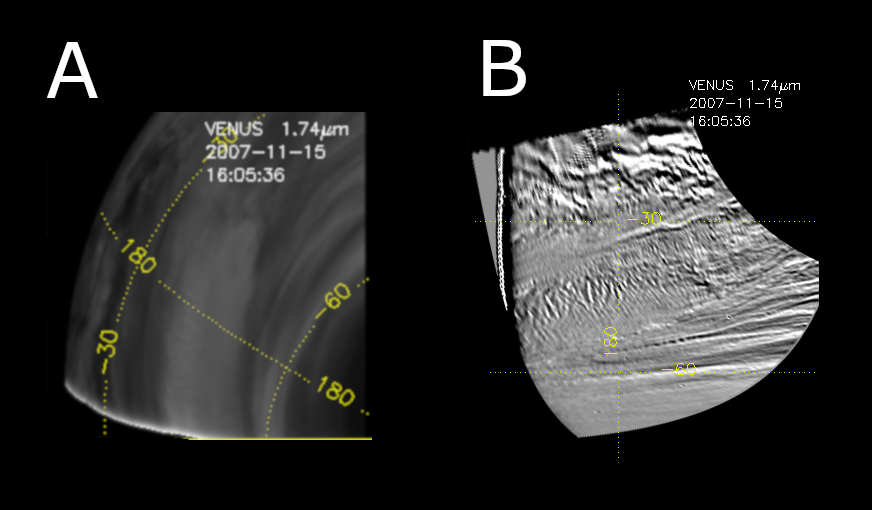}
                \caption{\footnotesize Example of a VIRTIS calibrated image from Venus Express archive (A) and after image processing and application of cylindrical projection (B).}
                \label{Process_Example}
        \end{figure}

                We performed a systematic search for periodic features on the lower cloud of Venus by visual inspection of each image. Each image was navigated and processed to increase the contrast on cloud features. The images were then mapped into cylindrical projections. Following previous observations of small-scale waves \citep{Peralta2008, Piccialli2014} we looked for wave packets with at least three bright and dark stripes such as those illustrated in Fig. \ref{wave_examples}. After images were navigated and image defects corrected, a number of contrast enhancement and unsharp mask filters were applied to the image so that the waves could be more easily recognised after processing, as can be seen in Fig. \ref{Process_Example}. Each positive detection was confirmed after checking the presence of the wave packet on images taken with different filters sensing the nightside lower clouds, eliminating spurious detections from additional defects. This was performed for the VIRTIS-M images. Confirmation through the identification of identical wave packets between images of the same area on the disc of Venus at short time intervals (1-2 hours) was also performed in some cases. 
        \subsection{Characterisation of wave properties}
                A total of 2685 VIRTIS-M cubes and 2878 IR2 images were inspected for the presence of atmospheric gravity waves. Positive detections of wave activity were characterised for their morphological (horizontal wavelength, packet length and width, orientation, and optical thickness drop) and dynamical (phase velocity and lifetime of wave packets) properties where possible. Processing techniques used for both detection and characterisation were similar to the ones described in \citet{Peralta2018} including limb fitting when necessary, as some nightside images showed a misplacement of the navigation grid. Adjustments in processing were always made on a case-by-case basis. VIRTIS images featured one exception as we found that applying a directional kernel for sharpening greatly improved the visibility of wave packets for those cases.\\
                Properties such as horizontal wavelength and packet width and length are extracted by calculating the distance between an origin and destination target points in the wave packet using the expression:\\
                \begin{equation}
                Dist = \frac{\pi \sqrt{(\lambda_2-\lambda_1)^2cos^2\left(\bar{\phi}\frac{\pi}{180}\right)+(\phi_2-\phi_1)^2}}{180}(a+h)
                \label{DistMeasurement}
                .\end{equation}
                This calculation is applied between two pixels on top of the visible disc of Venus (provided they are navigated). The values of $\lambda_1, \lambda_2$ and $\phi_1, \phi_2$ are the longitudinal and latitudinal coordinates of the origin and destination points, $\bar{\phi}$ is the average latitude between measured points, $a$ is the planet radius and $h$ is the altitude of the observed cloud layer.\\
                Horizontal wavelength and packet length and width measurements were performed several times for the same packet and then averaged. The orientation of the packet is the angle between the general axis perpendicular to the wave front alignment and the local parallel at the origin point. Angles in this calculation are in degrees.\\
                \begin{equation}
                \theta = arctan\left(\frac{\Delta \phi}{\Delta \lambda}\right)
                \label{Orientation_Eq}
                .\end{equation}
                Estimation of the wave amplitude is difficult because retrievals of atmospheric parameters like temperature or density from nadir observations of Venus's nightside lower clouds are subject to large uncertainties. However, the disturbance of waves  over the optical thickness of the  clouds can be used as a `proxy' for the normalised wave amplitude \citep{Tselioudis1992}. The calculation of the optical thickness perturbation by the wave packets is performed using the same formula as that described in \citet{Peralta2020}, taking the optical thickness drop ratio between crests and troughs. This calculation was not performed for wave packets in Akatsuki/IR2 images because of the problem of light pollution from the saturated dayside \citep{Satoh2017}. In VEx/VIRTIS images, wave packets close to the terminator were also excluded.\\[0.2cm]
                The dynamical characterisation of wave packets was more challenging. Due to the observational constraints previously mentioned, VIRTIS could not observe the same region of the disc continuously for more than 6 hours within the same orbit. This limited the number of wave packets that could be tracked individually to analyse their propagation and dynamical evolution.
                Nonetheless, a significant number of wave packets were effectively tracked, allowing us to record their phase speed and how they possibly interact with the background wind flow.\\[0.2cm]
                This task was better accomplished for waves detected with IR2 images because orbital aspects of the spacecraft permitted visualisation of larger areas over longer stretches of time than VIRTIS. This in turn allowed wave packets to be followed, on some occasions before they became apparent or after they vanished in the images.\\
                We used the same procedure as \citet{Peralta2018} to track the displacements of each crest of a wave packet between two images separated by a known time interval, retrieving at least ten wind tracers per packet. We also retrieved the local background wind at similar latitudes using methods identical to cloud tracking --- also applied in \citet{Peralta2018} and \citet{Goncalves2019} --- to evaluate the intrinsic phase velocity of these waves and how they relate to the general atmosphere dynamics.\\
                Phase velocities of wave packets and background wind velocity were measured individually using wind tracers and the following equation:
                \begin{equation}
                        U = \frac{cos(\bar{\phi})R_{eq}\frac{\pi}{180}\Delta Lon}{\Delta t}
                ,\end{equation}
                where U is the zonal velocity of the tracked tracer (wave crest for phase velocity or any cloud feature for the background wind), $\bar{\phi}$ is the latitude average as in equation \ref{DistMeasurement}, $R_{eq}$ is the equatorial radius of Venus, $\Delta$Lon is the zonal displacement of the tracked feature between images A and B, and $\Delta$t is the temporal interval between the considered images. Wave phase speed is measured by tracking individual structures that belong to the wave packet (mostly wave crests) between one or more pairs of images, depending on the lifetime and visibility of the packet. In addition to the above-mentioned constraints on the dynamical characterisation of waves, wave dispersion also plays a role in their visual classification and characteristics, as a single wave packet can break into two or more separate ones.
        \subsection{Theoretical considerations}
        \label{WaveTheory_Calc}
                Assuming that the wave packets apparent in the observations correspond to atmospheric gravity waves, a simple analytical model was employed for their interpretation. To this end, we made use of radio-occultation data from the Ultra-Stable Oscillator (USO) \citep{Imamura2011} onboard the Akatsuki space mission, namely temperature profiles to constrain the vertical wavelength of waves present at the nightside lower clouds. These temperature profiles were also used to infer the Brunt V\"{a}is\"{a}l\"{a} frequency, which can be used to unveil the true character of the waves present in observations.\\
                Using linear theory to describe wave phenomena, we can describe waves as a small perturbation of the mean flow on a specific atmospheric layer (in this case the lower clouds of Venus). Using an approach specific to slow terrestrial rotators such as Venus ---described in \citet{Peralta2014}--- and assuming that the vertical wind shear of the zonal wind does not significantly affect wave propagation, which can be verified if the intrinsic phase velocity is higher than the change in the zonal wind in a vertical wavelength \citep{IgaMatsuda2005}, and that the static stability is approximately constant within the altitude region studied, we can obtain a general dispersion relation for gravity waves in the form: 
                \begin{equation}
                        {(c_p^x - \bar{u})}^2 = \frac{N.k^2 + \xi^2.(m^2 + \frac{1}{4H^2})}{k^2 + m^2 + \frac{1}{4H^2}}
                        \label{Dispersion_Rel_cpx}
                ,\end{equation}
                where $c_p^x$ is the zonal component of the phase velocity, \={u} is the average zonal wind, N is the Brunt V\"ais\"{a}l\"{a} frequency, $\xi$ represents a centrifugal frequency modified by the meridional shear of the background wind, H is the density scale height, and k and m are the horizontal and vertical wave numbers. We are not including the meridional component of the phase velocity of waves in this discussion because the best spatial resolution achieved on these images is of the same order of magnitude as the error in meridional wind flow in the lower cloud, as already discussed by \citet{Hueso2012}. The Brunt V\"{a}is\"{a}l\"{a} frequency can be estimated using the results from radio occultation measurements performed by Akatsuki and described in detail in \citet{Ando2020}, namely the value of the static stability in the lower cloud of Venus during the period of observation:
                \begin{equation}
                        S = \frac{dT}{dz} + \frac{g}{C_p}
                ,\end{equation}
                where S is the static stability of the layer of the atmosphere, T and z are the temperature and altitude, g is the acceleration of gravity, and $C_p$ is the specific heat at constant pressure. From this, the Brunt V\"{a}is\"{a}l\"{a} frequency is calculated through:
                \begin{equation}
                        N = \sqrt{\frac{g~.~S}{T}}
                .\end{equation}
                As the static stability and consequently the Brunt V\"{a}is\"{a}l\"{a} frequency can both vary with altitude and latitude, values of N were calculated for different altitude levels within the expected altitude region of the  lower cloud. These analytical models are  compared to the retrieved data in dispersion diagrams in Sect. \ref{Discussion_NatureofWaves}.\\
                The estimation of the Brunt V\"{a}is\"{a}l\"{a} frequency is then used in the computation of the vertical wavelength of characterised packets along with the retrieved horizontal wavelengths and phase velocities:
                \begin{equation}
                        k = \frac{2\pi}{\lambda_x} \hspace{1cm} m = \frac{2\pi}{\lambda_z}
                ,\end{equation}
                where $\lambda_x$ and $\lambda_z$ are the horizontal and vertical wavelengths of packets. Introducing the latter in equation \ref{Dispersion_Rel_cpx} we are able to compute the vertical wavelength of wave packets with:
                \begin{equation}
                        \lambda_z = \frac{2\pi\sqrt{\hat{c_p^x} - \frac{\xi^2}{k^2}}}{\sqrt{N^2 - \hat{c_p^x}^2.k^2 + \frac{1}{4H^2}\left(\frac{\xi^2}{k^2} + \hat{c_p^x}^2\right)}}
                        \label{Vwave_eq}
                \end{equation}
                where $\hat{c_p^x}$ is the intrinsic zonal component of the phase speed resulting from $c_p^x$ - \={u}. We calculated the value of $\lambda_z$ for each wave packet at the altitudes within the lower cloud range, using the previously obtained values of the static stability from radio occultation data.\\
                For a more precise approach, we understand that both vertical shear of the zonal wind and variability of the static stability within the lower cloud layer exist which would necessarily change these equations. Even though we account for variations of the static stability with altitude when calculating $\lambda_z$, we address the implications of these approximations in Section \ref{Discussion_Vwave_Est}.
        
        \section{Results}
        
        Examples of wave packets observed and characterised with both VIRTIS and IR2 nightside images are displayed in Fig. \ref{wave_examples}. A total of 277 wave packets were identified and morphologically characterised (VIRTIS and IR2) while for dynamics only 168 characterisations were retrieved. Of all retrieved wave packets, approximately 32\% were dynamically characterised, with a significantly higher proportion of IR2 waves available for tracking than those from VIRTIS. Packets were characterised on every image in which they were present, not only for completion but also to track the evolution of each packet over time. As such, we distinguish every wave packet measurement with distinct packet measurements as the latter refers to a single wave packet propagating in the atmosphere, even if its properties are measured on more than one image. Regarding distinct packets, 94 were observed and analysed on VIRTIS images and 42 from IR2 data. With the addition of the data from \citet{Peralta2008}, a total of 166 different wave packets are included in this study.
        \subsection{Morphological properties}
        \begin{figure}[!]
                \centering\includegraphics[width=1.0\columnwidth]{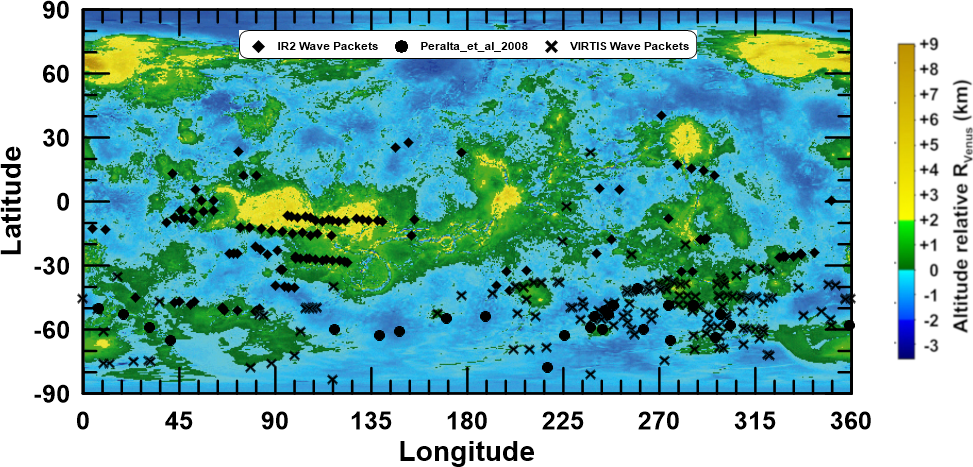}
                \caption{\footnotesize Distribution of characterised wave packets on the nightside of Venus during the period of observation. Wave packets from VIRTIS data are represented by crosses and from IR2 are represented by rhombuses. Additionally, represented by dark circles are wave packets featured in \citet{Peralta2008}. The topography map was made from data from VeRa onboard Venus Express \citep{Hausler2006, Hausler2007}.}
                \label{Waves_Distribution}
        \end{figure}
        \begin{figure}[!]
                \centering
                \includegraphics[width=1.0\columnwidth]{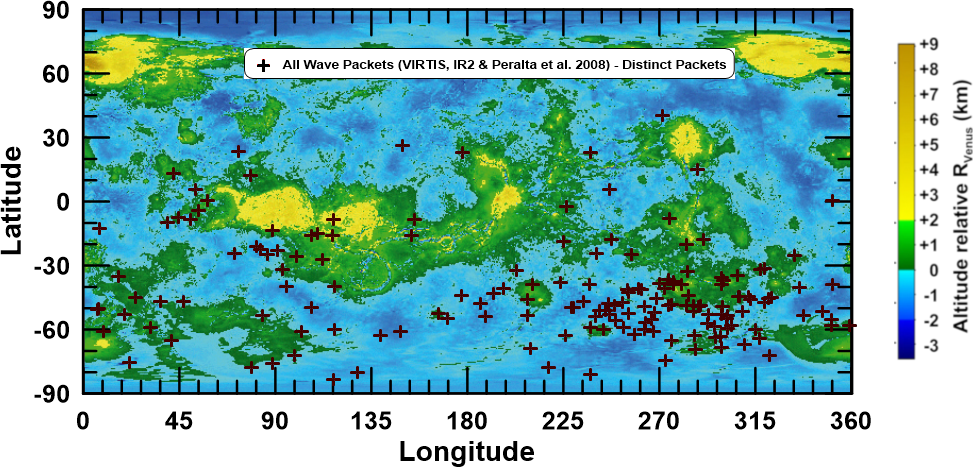}
                \caption{\footnotesize Distribution of characterised wave packets on the nightside of Venus in a latitude/longitude map. On this map, we represent only distinct packets from all three data sets: VIRTIS, IR2, and the data from \citet{Peralta2008}.\label{Waves_Distribution_Distinct}}
        \end{figure}
\begin{figure}
                \centering
                \includegraphics[width=0.9\columnwidth]{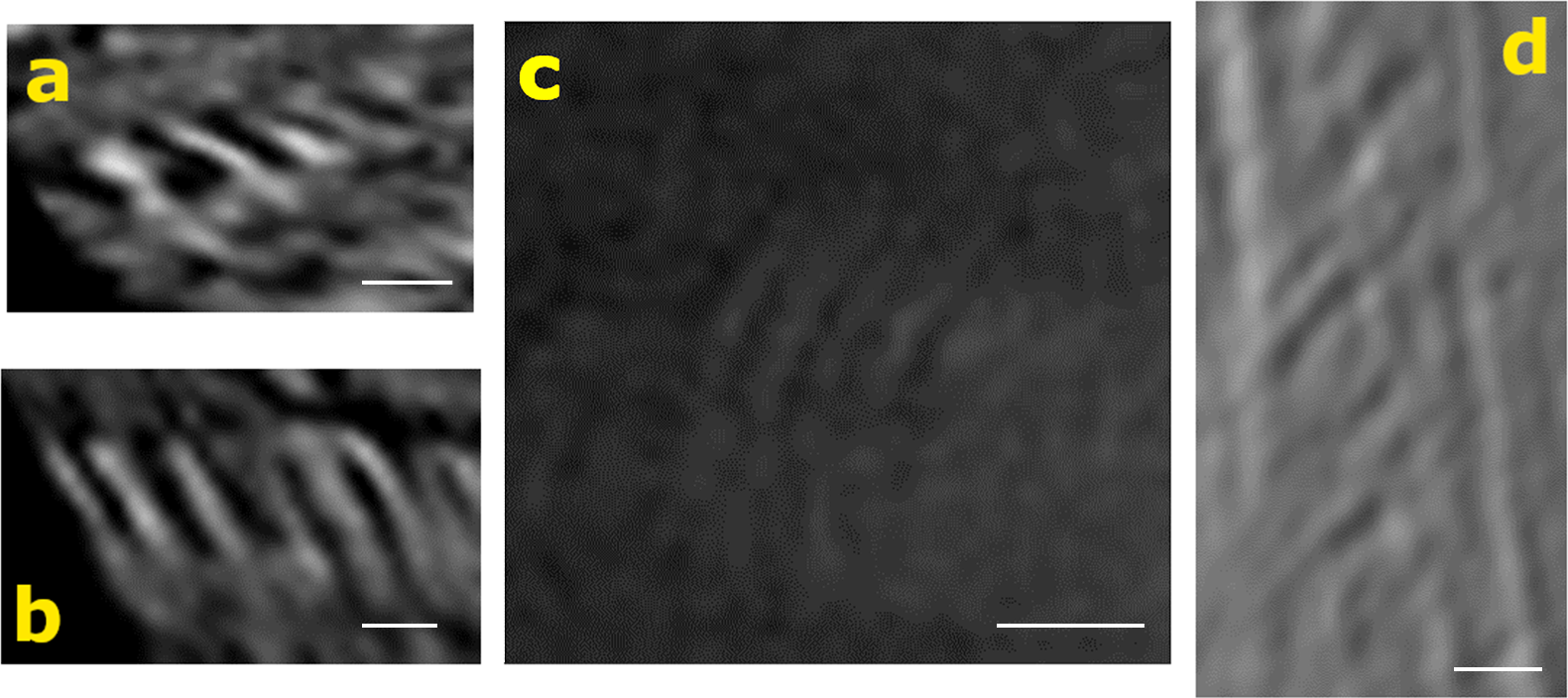}
                \caption{\footnotesize Examples of wave packets characterised with VIRTIS and IR2. The two images on the left (a(VI0834\_04), b(VI0607\_07)) show the crests and troughs very clearly, with a sharp contrast in comparison to the background atmosphere. The two images on the right (c(ir2\_20160905\_033333\_226\_l2b\_v10), d(VI0588\_05)) show that the boundary for the width of each different crest is not as clear. The white bars on each image represent a distance of 100 Km.\label{wave_Boundary}}
        \end{figure}

        Figure \ref{Waves_Distribution} shows every instance of wave-packet characterisation, even if the same wave packet is being characterised across different images. In Fig. \ref{Waves_Distribution_Distinct} we also present a condensed version in which we show only distinct packets. The large rows of points that represent the movement of the same wave packet characterised on different images are gone as a result. As illustrated by both Figs. \ref{Waves_Distribution} and \ref{Waves_Distribution_Distinct}, most of the wave activity was observed on the southern hemisphere of Venus.\\
        \begin{figure}[!]
                \centering
                \includegraphics[width=1.0\columnwidth]{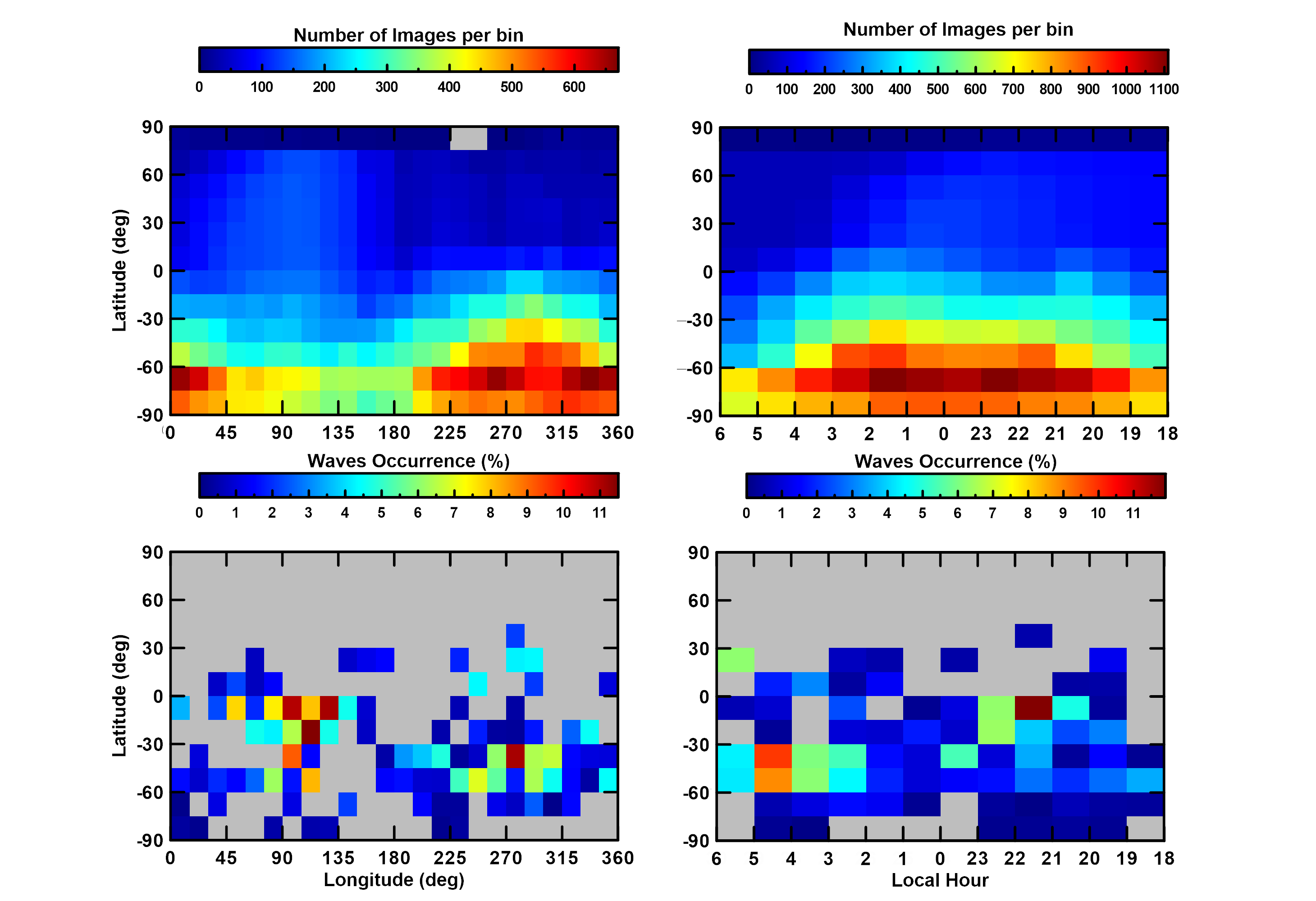}
                \caption{\footnotesize{\textit{Top} - Latitude/longitude and latitude/local time coverage maps of VIRTIS and IR2 images during the period of observation for both datasets. A greater number of images are shown for the southern hemisphere, particularly in regions between 60$^{\circ}$ and 90$^{\circ}$S at 0$^{\circ}$-45$^{\circ}$ and 195$^{\circ}$-360$^{\circ}$ and slightly decreasing between both terminators. \textit{Bottom} - Latitude/longitude and latitude/local time maps of the percentage of wave occurrence within the number of images analysed.}}.
                \label{NImages_Wave_Occurence_Panel}
        \end{figure}
        Moreover, the large concentration of VIRTIS packets at 225$^{\circ}$-315$^{\circ}$ longitude may be an observation bias, because during the observed period (August 2007 - October 2008) VIRTIS observed this region much more frequently than other areas which is illustrated in Fig. \ref{NImages_Wave_Occurence_Panel} (Top-left plot). There are also a good number of images and detections at equatorial and `subtropical' latitudes (0$^{\circ}$-30$^{\circ}$)  as the orbit of Akatsuki enables the detection of wave activity closer to the equator and on the northern hemisphere on Venus, which is not possible with VEx/VIRTIS.\\
        The occurrence maps in Fig. \ref{NImages_Wave_Occurence_Panel} combine the distribution of characterised waves with the number of images that target each sector on Venus to show the mesoscale wave frequency (number of observations with mesoscale waves to `total number of observations' ratio) at different locations. Even though wave occurrence is never higher than 11\%, an asymmetry in their distribution is clear as wave occurrence seems more concentrated in two different regions, between 45$^{\circ}$and 135$^{\circ}$ at equatorial latitudes and between 270$^{\circ}$and 315$^{\circ}$ flanked by subtropical latitudes and the region where the cold-collar would be \citep{Piccialli2012}.\\
        \begin{figure*}[t]
                \centering
                \includegraphics[width=0.6\paperwidth]{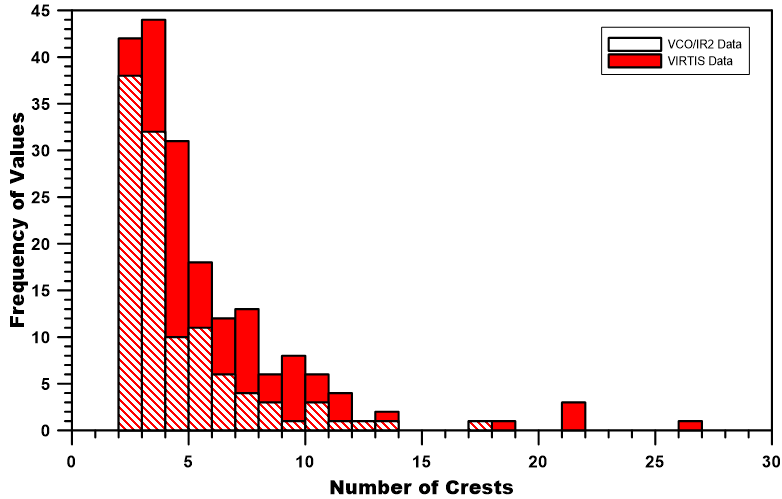}
                \caption{\footnotesize Histogram showing the number of crests on each characterised wave packet. Includes data from \citet{Peralta2008} which encompass VIRTIS data from July 2006 to March 2007.}
                \label{Hist_NCrests}
        \end{figure*}
        Figure \ref{Hist_NCrests} shows the number of crests of characterised wave packets. A minimum of three crests for wave packet detection was required to distinguish mesoscale waves from other cloud patterns that might share a similar morphology. There seems to be a higher occurrence of shorter wave packets (3-4 crests) in both data sets which decreases drastically for packets with many more crests (>10). However, the causes of these observations are uncertain.\\
        \begin{figure*}
                \centering
                \includegraphics[width=0.8\paperwidth]{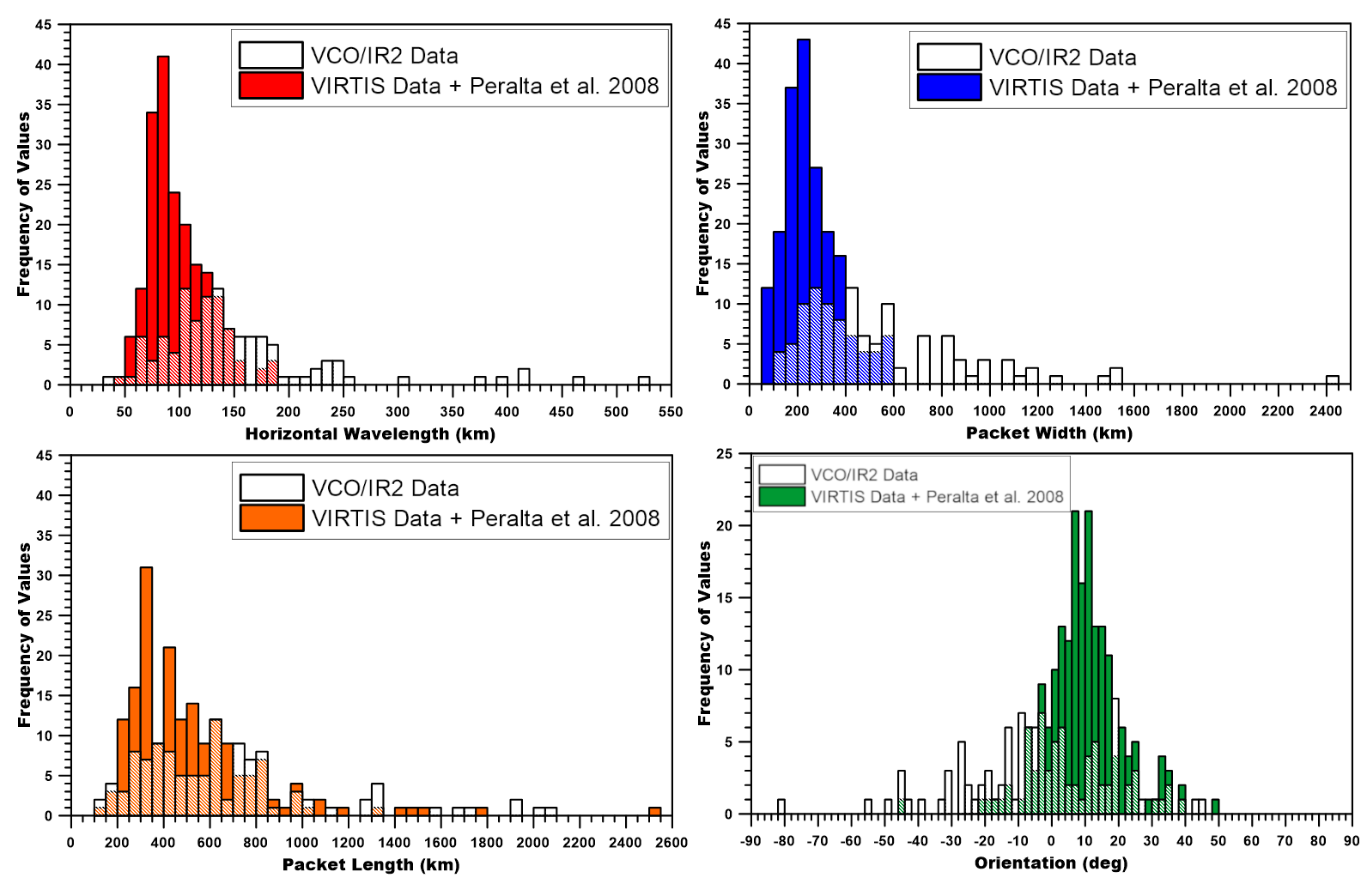}
                \caption{\footnotesize Histogram plots of the morphological properties of identified waves on nightside images of VIRTIS-IR and VCO-IR2. Also included are data from \citet{Peralta2008}.}
                \label{Hist_AllMorph}
        \end{figure*}
        Figure \ref{Hist_AllMorph} shows the results of the morphological properties of characterised packets. Waves characterised with VIRTIS data have the values of their properties, such as horizontal wavelength and packet width, more concentrated on narrower ranges than IR2 waves. Even though we have approximately 59\% more wave characterisations with VIRTIS data when compared with IR2, the properties distribution is not proportional between both data sets, especially for the case of the orientation of wave packets in Fig. \ref{Hist_AllMorph} (Bottom right plot) where we can identify approximately the 10$^{\circ}$ orientation as the most frequent value for VIRTIS waves, whereas IR2 waves present two peaks regarding orientation, one at -10$^{\circ}$and another at 20$^{\circ}$.\\
        \begin{table*}
                \centering\footnotesize
                \caption{Morphological properties of characterised packets}
                \begin{tabular}{ccccccccc}
                        \hline\hline
                        \noalign{\vspace*{0.2cm}}
                        Instrument & $\lambda_x$ & $\sigma_{\lambda_x}$ & PW & $\sigma_{PW}$ & PL & $\sigma_{PL}$ & $\theta$ & $\sigma_{\theta}$\\
                         & \multicolumn{6}{c}{(Km)} & \multicolumn{2}{c}{($^\circ$)}\\
                        \noalign{\vspace*{0.2cm}}
                        \hline
                        \noalign{\vspace*{0.2cm}}
                        VIRTIS & 100 (48 - 183)& 27.07 & 250 (77 - 597) & 115 & 486 (137 - 2512) & 302.77 & 9 (-45 - 50) & 11.61 \\[0.1cm]
                        IR2 & 158 (39 - 524) & 82.79 & 527 (115 - 2340) & 360.16 & 716 (107 - 2089) & 412.24 & -3 (-80 - 45) & 21.78 \\[0.2cm] 
                        \hline
                \end{tabular}
                \tablefoot{$\lambda_x$ is the horizontal wavelength, PW is the packet width, PL is packet length, and $\theta$ is the orientation of the packet. $\sigma_{\lambda_x}$, $\sigma_{PW}$, $\sigma_{PL}$ and $\sigma_{\theta}$ are the standard deviations between all wave packets for its respective property. The first value of each of $\lambda_x$, PW, PL, and $\theta$ is the mean value and in brackets are the minimum and maximum values measured for that property.}
                \label{MorphWavesTable}
        \end{table*}
        The values retrieved for the morphological properties of wave packets on both VIRTIS and IR2 databases are summarised in Table \ref{MorphWavesTable}. 
        The spatial resolution of observations was limited (minimum of 12 km/pix on VIRTIS and 5 km/pix on IR2) and so there is the possibility of waves with shorter wavelengths that we were unable to characterise in our observations. 
        Also, the widths of some of these packets changed considerably within their extension (packet length), which lead to larger deviations from the mean value of the packet width.
        Furthermore, some of the packets did not have a clear boundary from where the crests emerged and their width might be greater than what observing conditions (the contrast between crest and background) would allow us to see during characterisation (see Fig. \ref{wave_Boundary}).\\
        As IR2 data span wider areas on Venus's nightside, higher values for the packet length are expected. However, some of these characterisations offer only the minimum packet length, as putative crests blend into the background atmosphere or the packets extend beyond the image, especially for the VIRTIS case. As the orientation is defined as the angle relative to the parallel (line of constant latitude parallel to the equator), values are positive when increasing to the north and negative otherwise. Information on the direction of propagation can only be confidently retrieved when the same wave packet can be identified in two or more images. As previously stated, orientation values of VIRTIS and IR2 waves have the most contrasting distributions when compared with other morphological properties. Probable cause is attributed to the broader region where the packets are located when compared to VIRTIS waves, although an explanation as to why orientation seems more affected by this than other properties remains unclear.\\
        \begin{figure}
                \centering
                \includegraphics[width=1.0\columnwidth]{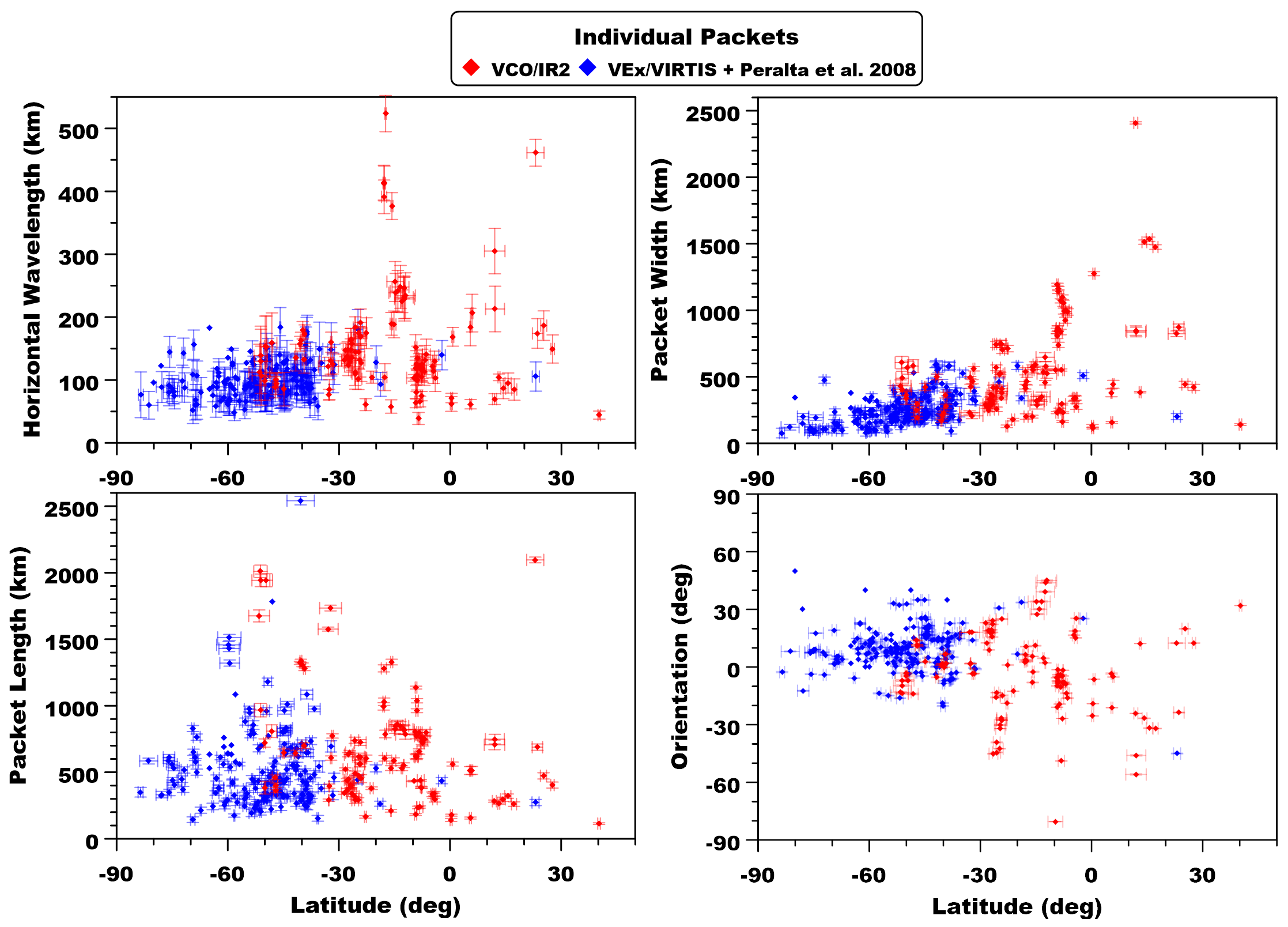}
                \caption{\footnotesize Distribution of morphological properties of waves with their respective latitude. Each plot point represents a wave packet detected with VIRTIS (blue or dark grey) or IR2 (red or light grey).}
                \label{Dist_Morph_vs_Lat}
        \end{figure}
        Figure \ref{Dist_Morph_vs_Lat} shows how the morphological properties of waves are distributed across the geographical latitude of Venus. Mesoscale waves at lower latitudes exhibit more scattered values for certain morphological properties, in contrast with waves at higher latitudes. Distinct sources and greater variability of the dynamics of the lower cloud layer at these latitudes might contribute to the dispersion of these values, creating more diverse waveforms \citep{Horinouchi2017}.\\
        \begin{figure}
                \centering
                \includegraphics[width = 0.8\columnwidth]{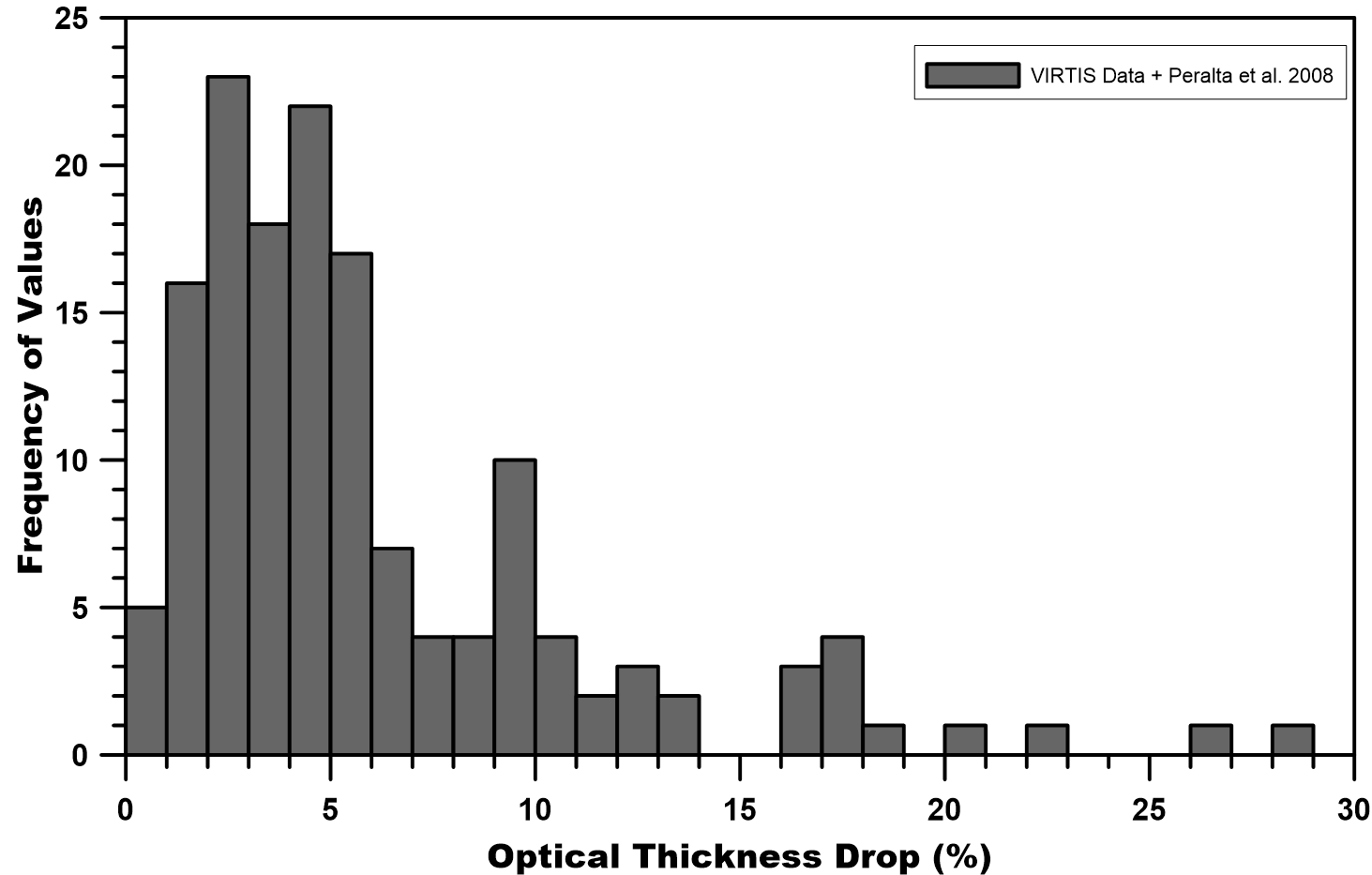}
                \caption{\footnotesize Histogram of the optical thickness drop ratio values for characterised packets on VIRTIS images.}
                \label{Hist_OptThick}
        \end{figure}
        Figure \ref{Hist_OptThick} shows the range of values for the relative drop in optical thickness between crests and troughs for characterised packets. We only include data retrieved from VIRTIS images because IR2 images are affected by light pollution as described in \citet{Satoh2017}. This aberration has a significant effect on the observed pixel information from which we would calculate the optical thickness, in most cases making it difficult to accurately measure the drop in optical thickness between crests and troughs with confidence.\\ 
        Data from VIRTIS images show that for most packets the relative drop is fairly low, namely between 2\% and 6\%, with packets whose relative optical thickness drops by higher values (> 10\%) being increasingly less frequent. However, we did not find any relation between this sharp decrease in optical thickness and any other properties or location of the wave packets. Nevertheless, there have been investigations into the relationship between optical properties of clouds and their temperature \citep{Tselioudis1992, Tselioudis1994}. It might be possible to use radiative transfer models to infer other properties of waves such as amplitude.
        \subsection{Dynamical properties}
        \label{Characterisation_DynamicalProp}
        With techniques akin to cloud tracking described in \citet{Machado2017}, \citet{Peralta2018} and \citet{Goncalves2019}, we measured the phase speed of 50 different wave packets on both instruments.
        \begin{figure}[!]
                \centering
                \includegraphics[width = 0.9 \columnwidth]{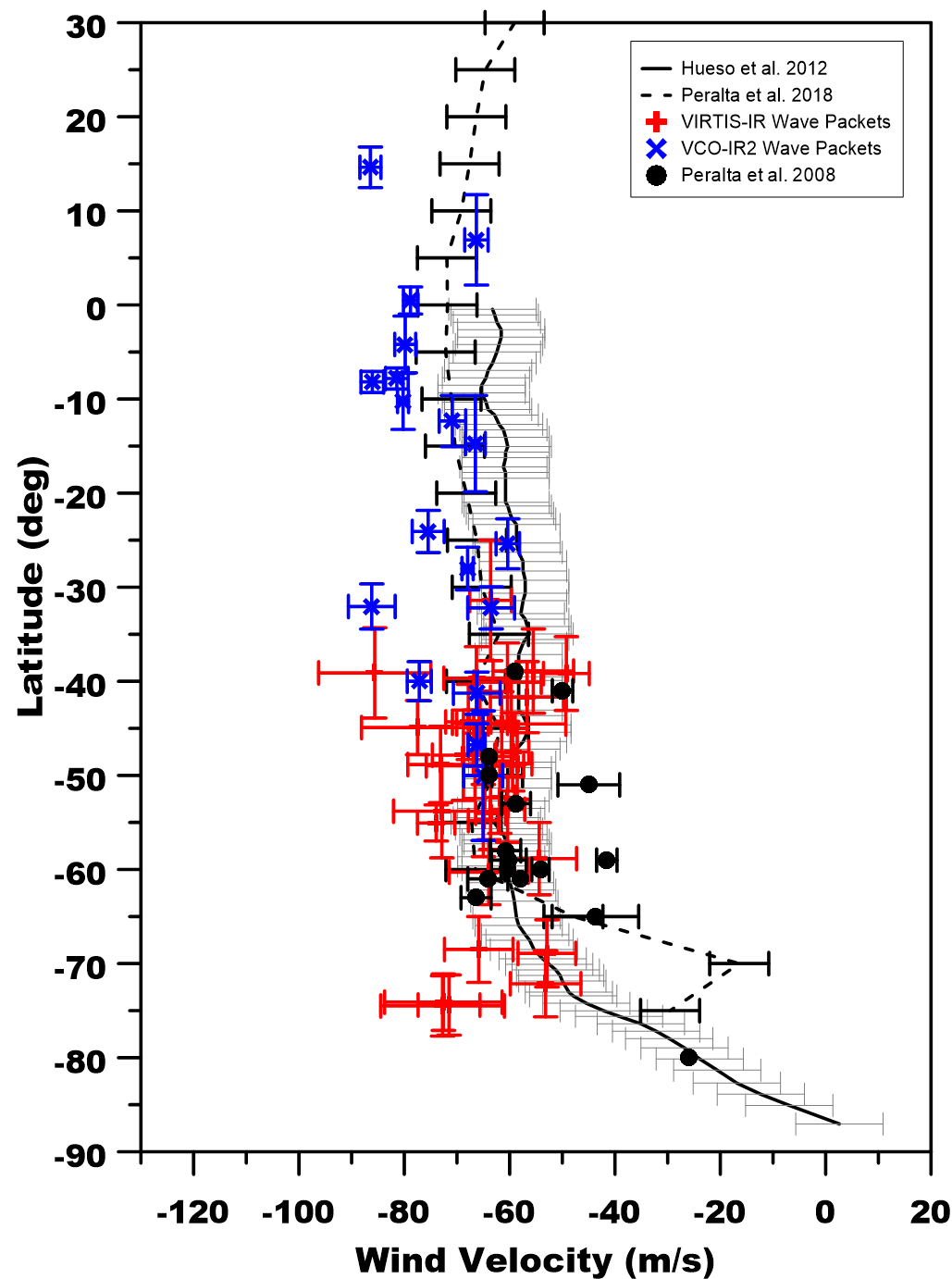}
                \caption{\footnotesize Zonal wind profile of Venus' lower cloud and the measured phase velocity of characterised packets from VIRTIS and IR2 data. The filled and dashed profiles in black represent the wind profiles on the lower cloud reported in \citet{Hueso2012} and \citet{Peralta2018}, respectively. Blue crosses (dark grey) and red plus signs (light grey) mark the absolute phase velocity of wave packets retrieved with IR2 and VIRTIS respectively along with data from \citet{Peralta2008}.}
                \label{ZonalWindProfile_Waves}
        \end{figure}
        Figure \ref{ZonalWindProfile_Waves} shows the measured wave-packet phase velocity compared with the mean zonal wind profiles for the lower cloud of Venus, where these waves were detected. Each phase velocity data point represents a different packet.\\
        \begin{figure}
                \centering
                \includegraphics[width = 0.9\columnwidth]{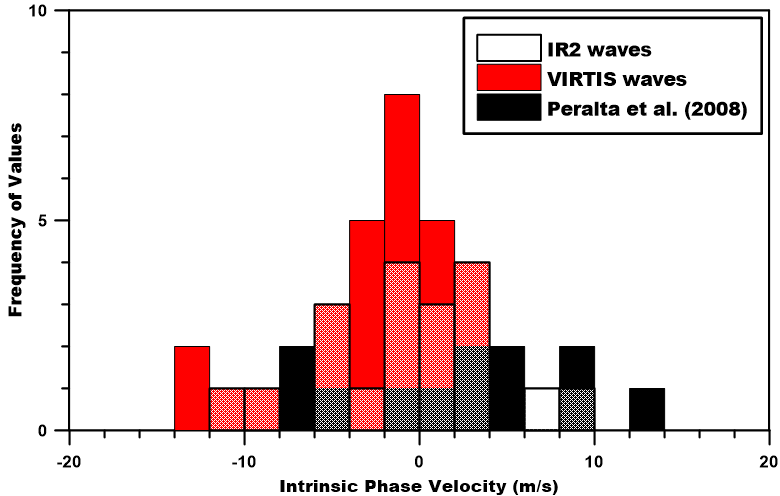}
                \caption{\footnotesize Histogram of intrinsic phase velocity of characterised packets. Bin size is 2 m/s and each value represents a different packet. Red (light grey) bins show values from VIRTIS data and overlaid on top with a semi-transparent pattern are values from IR2 data. Additionally, as black bins we show the data from \citet{Peralta2008}.}
                \label{Hist_IPV}
        \end{figure}
        Figure \ref{Hist_IPV} shows the intrinsic phase velocity of measured waves. These are the values presented in Fig. \ref{ZonalWindProfile_Waves} with the retrieved local background wind subtracted. As this calculation is performed as $\hat{c_p^x}$ = $c_p^x$ - \={u}, where $\hat{c_p^x}$ is intrinsic phase velocity, $c_p^x$ is the measured phase velocity, and \={u} the local background wind, along with Venus featuring a retrograde wind flow, negative values imply a wave with its phase speed faster than the local wind and, conversely, positive values imply slower wave packets relative to the background wind.\\
        \begin{figure}
                \centering
                \includegraphics[width = 1.0 \columnwidth]{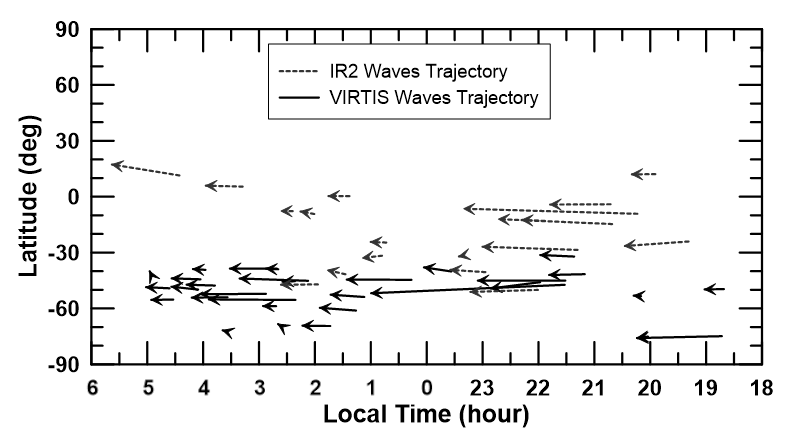}
                \caption{\footnotesize Trajectory of tracked wave packets from VIRTIS (filled line) and IR2 (dashed line) data on a latitude/local time map of the nightside of Venus. The length of the arrows represents the location of the first observation of the packet, following a straight trajectory to its last observed location.}
                \label{Lifetime_LT_VIRTISIR2}
        \end{figure}
        \begin{figure}
                \centering
                \includegraphics[width = 1.0 \columnwidth]{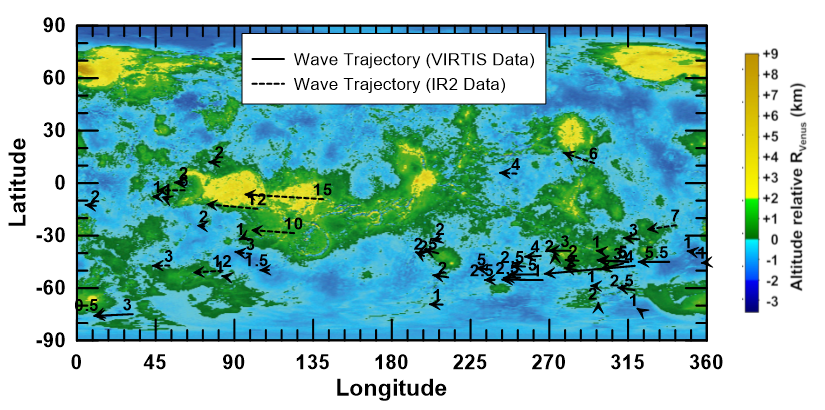}
                \caption{\footnotesize Trajectory of tracked wave packets from VIRTIS (filled lines) and IR2 (dashed lines) data from the location of the first characterisation to the final location. The labels on each arrow represent the minimum lifetime in hours of the respective packet. Topography map from Magellan probe data in the background.}
                \label{Lifetime_VIRTISIR2}
        \end{figure}
        Figures \ref{Lifetime_LT_VIRTISIR2} and \ref{Lifetime_VIRTISIR2} show how wave packets propagate during observation. The latitude/longitude map shows the travelled distance in a straight line between the first tracked position to its last known location. Initial positions are not necessarily related to the formation of the wave packet, nor is the final position linked with the dispersion of the wave. The labels on each arrow represent the minimum lifetime of tracked waves as they propagate on the atmosphere of Venus. As already demonstrated in Fig. \ref{Hist_AllMorph}\textcolor{webgreen}{.D}, most of the packets follow the dominant zonal wind flow along their trajectory which is further illustrated here as well as in Figs. \ref{Lifetime_LT_VIRTISIR2} and \ref{Lifetime_VIRTISIR2}.\\
        \begin{figure}
                \centering
                \includegraphics[width=1.0\columnwidth]{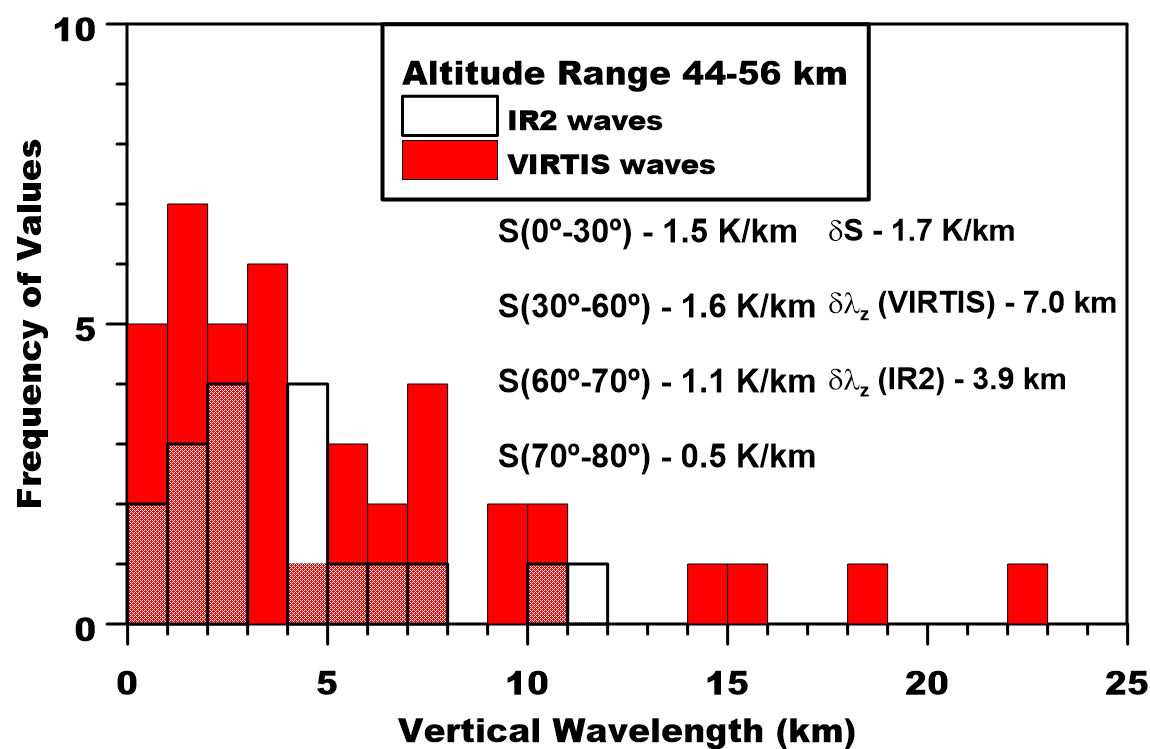}
                \caption{\footnotesize Histogram of the vertical wavelength of distinct characterised packets for the altitude range (44-56 km). We also present here the values of the static stability used to compute the vertical wavelengths for each latitude range (S), the mean error on the static stability ($\delta$S) from \citet{Ando2020}, and the propagated errors for the vertical wavelength for both instruments ($\delta\lambda_z$). The bins are 1 km wide.}
                \label{Hist_Vwave}
        \end{figure}
        Figure \ref{Hist_Vwave} presents a histogram plot of the vertical wavelengths calculated with the dispersion relation from Sect.\ref{WaveTheory_Calc} (Eq. \ref{Vwave_eq}). As the static stability varies considerably with latitude within the lower cloud layer, we calculated the vertical wavelengths using four different static stability values shown in Fig.\ref{Hist_Vwave} which were determined from temperature profiles retrieved from radio occultation data in \citet{Ando2020}. We considered a larger altitude range of $\sim$ 44-56 km to calculate the values for $\lambda_z$ in order to accommodate the possible variability of the altitude range where the lower cloud is estimated to be \citep{Titov2018}. 
        \subsection{Characterisation precision}
        The spatial resolution of images is the most important aspect concerning both atmospheric wave detection and their morphological
and dynamical characterisation. In turn, these are highly dependent on the proximity of the spacecraft to the target location being monitored. Given the orbital characteristics of both spacecraft detailed in \citet{Svedhem2007} for Venus Express and \citet{Nakamura2016} for Akatsuki, the mean spatial resolution obtained on images with characterised waves was $\sim$ 23.9 km/pix and $\sim$ 20.6 km/pix for VIRTIS and IR2 data, respectively, with better resolution towards higher latitudes for the former and closer to the equator for the latter.\\
        \begin{table*}
                \centering\footnotesize
                \caption{Morphological characterisation precision}
                \begin{tabular}{cccccc}
                        \hline\hline
                        \noalign{\vspace*{0.2cm}}
                        Instrument & Spatial resolution & $\zeta_{\lambda_x}$ & $\zeta_{PW}$ & $\zeta_{PL}$ & $\zeta_{\theta}$\\
                        & (km/pix) &\multicolumn{3}{c}{(Km)} & ($^{\circ}$)\\
                        \noalign{\vspace*{0.2cm}}
                        \hline
                        \noalign{\vspace*{0.2cm}}
                        VIRTIS & 23.9 (12.78 - 42.6)& 12.23 (12.23\%) & 24.2 (9.68\%) & 18.63 (3.83\%) & 1.54 (17.11\%)\\[0.1cm]
                        IR2 & 20.6 (5.32 - 51.19)& 21.95 (13.89\%) & 51.98 (9.86\%) & 36.92 (5.16\%) & 2.3 (76.67\%) \\[0.2cm] 
                        \hline
                \end{tabular}
                \tablefoot{Mean spatial resolution values followed by their respective minimum and maximum values and individual measurement errors for each morphological property measured. The percentage value is relative to the mean values retrieved for each property in Table \ref{MorphWavesTable}.}
                \label{MorphErrorsTable}
        \end{table*}
        The contents on Table \ref{MorphErrorsTable} show the measurement error for each morphological property which is given directly by the spatial resolution of the image where a wave packet is detected. As discussed in Sect. \ref{DataAcquisition} this resolution is highly dependent on the distance between the spacecraft and the target, controlled by its orbit around Venus. The values for $\zeta_{\lambda_x}$, $\zeta_{PW}$, $\zeta_{PL}$, and $\zeta_{\theta}$ represent the mean standard deviation between measurements of the same wave packet for each respective property. With these values, it is possible to gauge the consistency of the measuring process within each wave packet. As the error in the distance measurement calculation (see eq.\ref{DistMeasurement}) is the same between measuring $\lambda_x$, PW, or PL, and as the two latter values are usually higher than the former, we expect a decrease in percentage error even if the absolute values for $\zeta_{\lambda_x}$, $\zeta_{PW}$, and $\zeta_{PL}$ rise beyond the mean spatial resolution.\\
        The value for $\zeta_{\theta}$ in Table \ref{MorphErrorsTable} is remarkably higher in percentage because the mean value for orientation of packets approaches 0$^{\circ}$ (see Table \ref{MorphWavesTable}). The precision of the orientation values also varied with the length of the wave packet. Due to the measuring process as well as the limited spatial resolution of the images, it was possible to determine the orientation of longer wave packets   more accurately than it was for shorter wave packets. The measuring process for orientation was also sensitive to the packet width, although to a lesser degree.\\
        Retrieved phase velocities are affected by the spatial resolution of each image and the time interval between them. The error in the velocity retrieval is calculated as:
        \begin{equation}
        \delta U = \frac{\delta s}{\Delta t}
        \label{Vel_Error_eq}
        ,\end{equation}
        where $\delta U$ is the velocity error, $\delta$s is the spatial resolution of the images (in meters/pixel), and $\Delta$t is the time interval between images. Hence, larger time intervals lead to smaller errors. The average error for measured phase velocities and background wind is $\sim$ 6.2 m/s within the range 2.5 - 12 m/s for VIRTIS waves and $\sim$ 2.2 m/s within 1 - 5 m/s for IR2 waves.\\
        Other predominant effects that compromise characterisation, especially optical thickness studies, is the light pollution from the dayside caused by multiple reflections of light inside the detector of IR2 \citep{Satoh2017}. Not only did this influence detectability of crests closer to the terminator (or possibly entire wave packets) but it also made some optical thickness characterisation impossible, as this external light would eclipse the natural drop between crests of a wave. For this reason, we elected to discard the optical thickness measurements from IR2 images. 
        Other sources of error include navigation and geometry errors from limb fitting, however these are generally less significant. Navigation of VIRTIS and IR2 images comes mainly from Spacecraft Planet Instrument C-matrix Events (SPICE) data and its main sources of error are considered to be the accuracy and stability of the spacecraft's attitude, the uncertainty in the spacecraft's position, and the time accuracy on the time-tag attributed to each image. For the case of VEx/VIRTIS, the spacecraft's attitude is usually stable, ensuring any deviation is not larger than $10$ mdeg, which is more than five times smaller than the best spatial resolution of the data used. The uncertainty in the position of the spacecraft leads to errors on the order of tens of meters, as analysed in \citet{Rosenblatt2008}, which is even lower and the time accuracy when tagging image data is on the order of milliseconds, more than five orders of magnitude smaller than the shortest time interval between images with waves.\\
        Navigation of IR2 images required more corrections because for many nightside images, light contamination from the dayside mentioned earlier along with other effects such as the high opacity of the lower clouds, which makes the limb more difficult to identify, led to significant misalignments of the navigation grid. As such, we used the same interactive limb-fitting method developed and detailed in \citet{Peralta2018}. The precision of this correction can be estimated to range between 0.5 and 1 pixel, depending on the case at hand.
        
        \section{Discussion}
        \subsection{The nature of characterised waves}
        \label{Discussion_NatureofWaves}
        Characterised waves are interpreted as internal gravity waves due to their neighbouring conditions (static stability) as well as their combination of characteristics which are not consistent with other types of waves that could form in Venus' atmosphere \citep{Peralta2008}. To further support this interpretation, we place the retrieved values on a dispersion diagram, which connects the type of wave propagating with its horizontal wavelength and intrinsic phase speed.\\
        \begin{figure}[h]
                \centering
                \includegraphics[width=1.0\columnwidth]{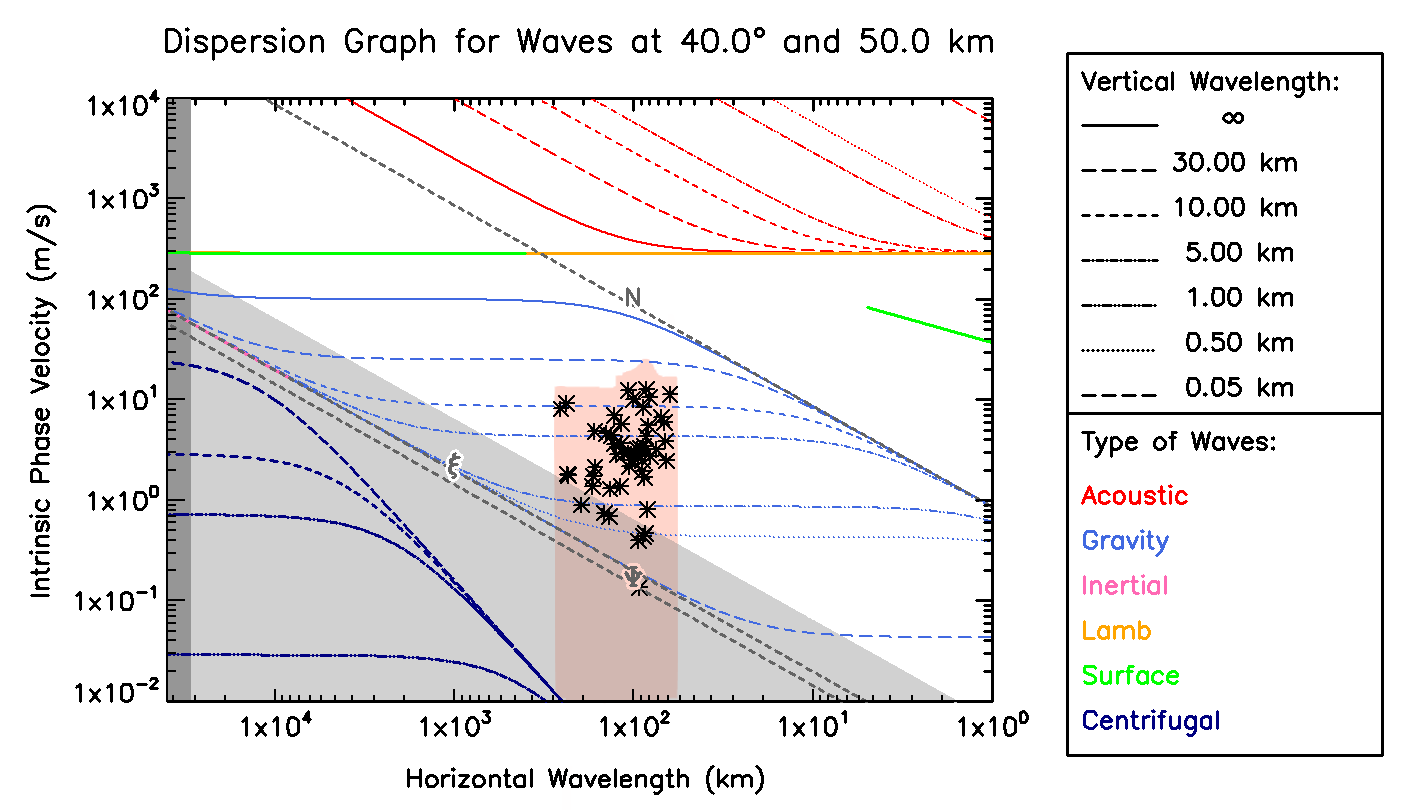}
                \caption{\footnotesize Dispersion diagram for dynamically characterised waves on both data sets. Each dashed coloured line represents the value of the vertical wavelength that a wave would have given its specific horizontal wavelength and intrinsic phase velocity values according to the models described in \citet{Peralta2014,Peralta2014b}. The shaded region over the data points represents the error on the phase velocity of the waves. Due to the logarithmic nature of this diagram, the error bars go all the way down towards the abscissa.}
                \label{DispersionGraph}
        \end{figure}
        Figure \ref{DispersionGraph} shows theoretical models for waves of various types given their respective horizontal wavelengths and intrinsic phase velocities \citep{Peralta2014, Peralta2014b}. Reference values used to build these models are detailed in \citet{Seiff1985} and \citet{Taylor1985}, some of which are included in a compendium of models for the atmosphere of Venus called the Venus International Reference Atmosphere (VIRA). However, these dispersion diagrams also use more recent data from instruments onboard Venus Express for more robust modelling of the atmosphere \citep{Piccialli2010, Hueso2012}. Each type of coloured line (a combination of dashes and dots) represents the analytical solution for pure waves of the appropriate type with specific vertical wavelengths. Also represented with grey lines are the Brunt V\"{a}is\"{a}l\"{a} frequency (N), the centrifugal frequency ($\Psi$), and the centrifugal frequency modified by the meridional shear of the background zonal wind ($\xi$) \citep{Peralta2014, Peralta2014b}. Lastly, the dark and light grey shaded areas mark the limits of the maximum horizontal wavelength allowed at the respective latitude and where the usual condition in which the intrinsic frequency of waves is much greater than $\xi$ no longer applies, respectively. In summary, these three quantities (N, $\Psi$ and $\xi$) can serve as boundaries where internal atmospheric gravity waves manifest themselves and are able to propagate as such. Even though the shape of the dispersion diagram, particularly the values of N, $\Psi,$ and $\xi,$ are variable with latitude and altitude, we present a single dispersion diagram comprising all the data, having verified that for each of their respective conditions most characterised packets remain within the gravity wave region and that the error bars related to the measurements are large enough to justify the use of a single mean value for latitude and altitude to build the dispersion diagram in Fig.\ref{DispersionGraph}. More details on these models can be found in \citet{Peralta2014, Peralta2014b}.\\
        The position of most characterised wave packets (black crosses in Fig. \ref{DispersionGraph}) is well within the gravity wave region. A few of these packets fall under the light grey area of the plot where some assumptions regarding the dispersion relation used in this model are no longer valid, which makes these particular packets more difficult to interpret regarding their nature. However, considering the error bars present in Figs. \ref{DispersionGraph}, all packets can be considered as atmospheric gravity waves. 
        Their position in the dispersion diagram also provides an estimation of the vertical wavelength of wave packets represented by the dashed lines. The static stability used in the models, which enables the calculation of vertical wavelengths, comes from Venus Express Data and its calculation is described in \citet{Piccialli2010} and the lowest altitude the models reach is 50 km. 
        However, in this work we estimated the vertical wavelengths from equation \ref{Vwave_eq} and used updated values of temperature from radio-occultation profiles from Akatsuki \citep{Ando2020}.
        \subsection{Vertical wavelength estimation and altitude of waves}
        \label{Discussion_Vwave_Est}
        The values for the vertical wavelength are calculated for a wider region than what is estimated to be the lower cloud of Venus (44-49 km of altitude). According to \citet{Titov2018}, the altitude level for the lower cloud starts at approximately 47 km, stating that the boundary between the lower and middle cloud is not well defined, going as high as 56 km. We also know from several models of the atmosphere of Venus, such as those described in \citet{Lefevre2018} to generate waves through convection, that there is a highly convective zone above 50-51 km which makes propagation of gravity waves more difficult. Furthermore, gravity waves cannot be observed in a region where the static stability is zero \citep{Sutherland2010}. We chose the larger interval, which accommodates both interpretations, as we are calculating an average result for all these altitude levels and the mean altitude value coincides exactly with the region where waves should start propagating (on the region where the static stability approaches zero, close to 50 km in altitude). Additionally, the vertical extension of these waves should not be larger than the area on which they are propagating. According to Fig. \ref{Hist_Vwave} we have a substantial number of waves that extend beyond 5 km (larger than the 44-49 km altitude range), and so a larger altitude interval must be considered. Regarding the convection region where we should not see gravity waves, there is a possibility as discussed in \citet{Lefevre2020} of transmission of gravity waves through these impossible zones, much like a quantum tunnelling effect, to upper layers of the atmosphere of Venus, possibly depositing momentum and feeding the super-rotation of the upper clouds. However, we are probably seeing waves generated by the convective region between the lower and upper clouds that propagate downwards, and as the region below the supposed lower cloud (< 44 km) is stable until roughly 30 km, it is possible to have waves that vertically extend down to these levels. One possible way to distinguish the altitude range within which we see waves, and whether these are propagating upwards or downwards, could be via the verification of upper cloud images at the same geographical locations as the waves in this study and measure cloud properties and dynamics for any possible alteration due to the waves propagating in the lower cloud.\\
        Estimated values of the vertical wavelength do not take into account the presence of vertical wind shear of the zonal wind or that the static stability below the cloud layer (where it can propagate) changes with altitude, which influences the form of Eq. \ref{Vwave_eq}. Considering the effects of wind shear on the propagation of waves it is possible to verify whether or not the influence from vertical shear is great enough to produce significant changes to the vertical wavelength of characterised waves. We can use the relation $|\hat{c_p^x}| > \lambda_z.|\frac{\partial\bar{u}}{\partial z}$| from \citet{IgaMatsuda2005} to determine whether or not the waves studied here are fast enough to avoid perturbation by vertical wind shear within one vertical wavelength. We took this analysis to the dynamically characterised packets and concluded that all waves are indeed fast enough for the vertical shear of the zonal wind to be insignificant. To further develop our analysis of this issue we recalculated the vertical wavelength using a more complete equation which includes wind shear and performs the Wentzel-Kramers-Brillouin (WKB) approximation so that an analytical formula can be obtained to compute the vertical wave number. Further details on this equation and the approximations used can be found in the textbook by \citet{Nappo2002}:
                \begin{equation}
                        m^2(z) = \frac{N^2}{\hat{c_p^x}^2} + \frac{\partial^2\bar{u}}{\partial z^2}\frac{1}{\hat{c_p^x}}-\frac{1}{H \hat{c_p^x}}\frac{\partial\bar{u}}{\partial z} - \frac{1}{4H^2} - k^2
                        \label{Nappo_eq_Vwave}
                ,\end{equation}
        where the partial derivative terms of the background zonal wind correspond to the vertical shear and its variability within the middle-lower cloud. Vertical profiles of the zonal wind can be found in \citet{Peralta2014} which are based on data from Pioneer Venus probes as well as more recent cloud-tracking data of the upper and lower clouds.
        \begin{table*}[!]
                \centering\footnotesize
                \caption{Variable values on the estimation of $\lambda_z$}
                \begin{tabular}{lcccc}
                        \hline\hline
                        \noalign{\vspace*{0.2cm}}
                        Latitude Range & 0-30$^{\circ}$ & 30-60$^{\circ}$ & 60-70$^{\circ}$ & 70-80$^{\circ}$ \\
                        \noalign{\vspace*{0.2cm}}
                        \hline
                        \noalign{\vspace*{0.2cm}}
                        N $\times$ 10$^{-3}$ (s$^{-1}$) & (5.5 $\pm$ 5.2) & (5.7 $\pm$ 5.3) & (4.9$\pm$7.6) & (3.6$\pm$6.7) \\[0.1cm]
                        $\hat{c_p^x}$ (m/s) & \multicolumn{4}{c}{[-12.6 -- 12.7] $\pm$ 4.4 } \\[0.1cm]
                        $\frac{\partial\bar{u}}{\partial z} \times 10^{-3}$ (m.s$^{-1}$.Km$^{-1}$) & \multicolumn{4}{c}{-0.9 $\pm$ 2.3} \\[0.1cm]
                        $\frac{\partial^2\bar{u}}{\partial z^2} \times 10^{-6}$ (m.s$^{-1}$.Km$^{-1}$)& \multicolumn{4}{c}{1.2 $\pm$ 2.4}\\[0.1cm]
                        H (m) & \multicolumn{4}{c}{6380 \tablefootmark{a}} \\[0.1cm]
                        k $\times$ 10$^{-5}$(m$^{-1}$) & \multicolumn{4}{c}{[2.7 -- 10] $\pm$ 1.2} \\
                        \noalign{\vspace*{0.2cm}}
                        \hline
                \end{tabular}
                \tablefoot{The latitude range intervals are the same as in \citet{Ando2020}. Values for N are obtained using the equations in Sect.\ref{WaveTheory_Calc} and the temperature profiles obtained from radio occultation data from Akatasuki, described in \citet{Ando2020}. We present the value and its propagated error for each entry.\\
                For $\hat{c_p^x}$ and k we present its range of values (minimum and maximum) across both VIRTIS and IR2 datasets and the mean velocity error and propagated error for the wave number across all measurements.\\
                For the wind shear and its variability we present the mean within the largest altitude range described earlier (44-56 km) along with its standard deviation across all latitude ranges as all values are within the presented error.\\ \tablefoottext{a}{Scale-height reference value for the cloud layer of Venus. \citep{Peralta2014}}}
        \label{Vwave_CompleteEq_variables}
        \end{table*}
        In Table \ref{Vwave_CompleteEq_variables} we present the range of values considered for the different variables at play in Eq.\ref{Nappo_eq_Vwave}. Wind shear values are obtained from the vertical profile of the zonal wind in \citet{Peralta2014}.\\
        Comparing the results between estimated vertical wavelengths obtained from Eqs.\ref{Vwave_eq} and \ref{Nappo_eq_Vwave} we arrive at an approximately 4\% relative difference between the results from both equations. These results, which are consistent with the verification from \citet{IgaMatsuda2005}, allow us to conclude that the vertical shear, although present on the sounded region of the atmosphere, has insufficient influence to perturb the vertical extension of the characterised waves.
        \subsection{Dependence on latitude/local time and influence on wave properties}
        Figures \ref{Lifetime_LT_VIRTISIR2} and \ref{Lifetime_VIRTISIR2} show that in general the waves identified in this work mostly follow a zonal downstream propagation and as we see in Fig. \ref{Hist_AllMorph}\textcolor{webgreen}{.D}, the wave fronts are generally perpendicular to parallel lines moving the perturbation mostly westward. As gravity waves seem partially limited by $\xi$ in terms of propagation \citep{Peralta2014}, we investigated the relationship between morphological aspects of waves and the retrieved dynamic properties.\\[0.2cm]
        \begin{figure}[!]
                \centering\includegraphics[width = 0.9\columnwidth]{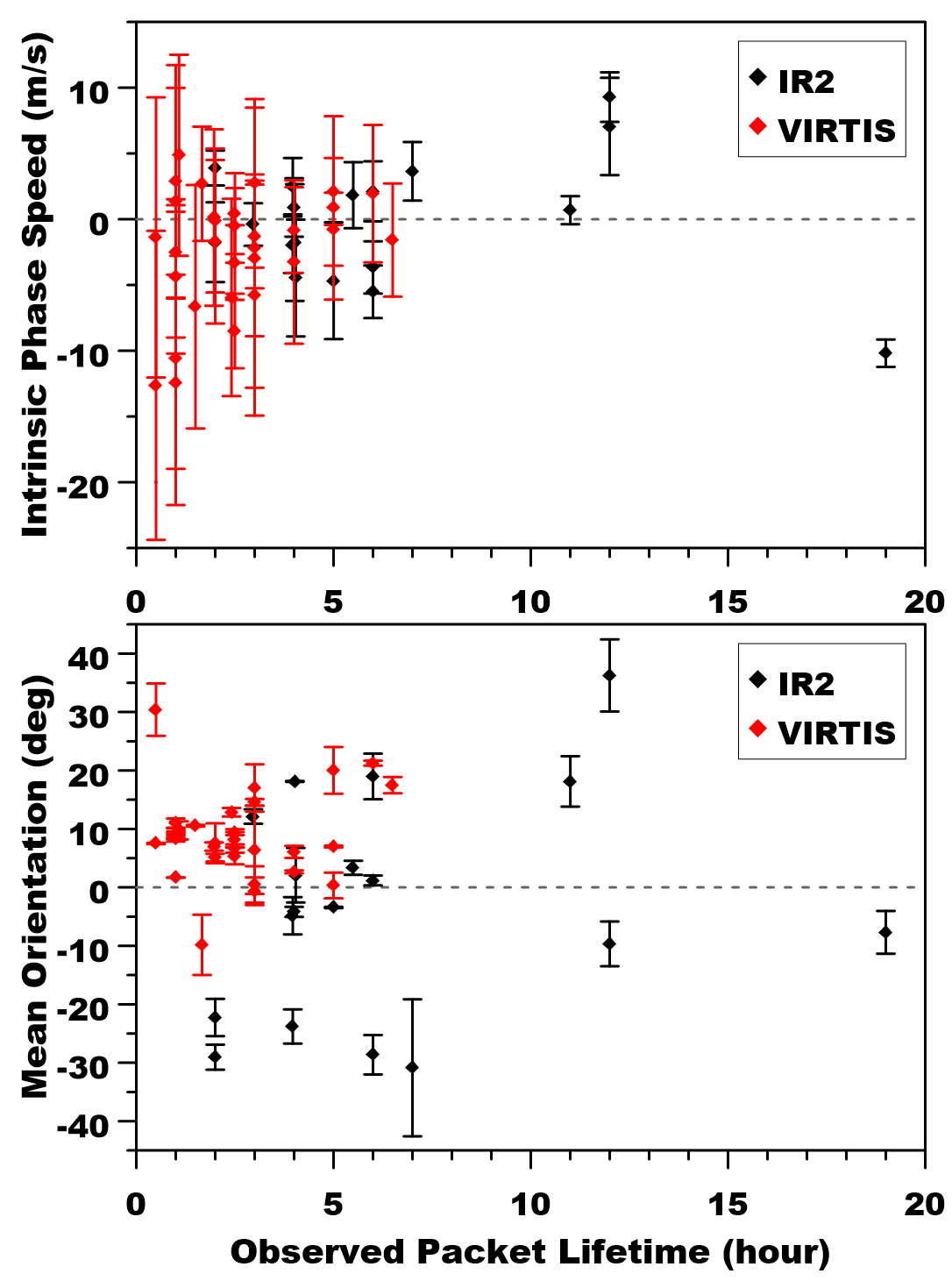}
                \caption{\footnotesize Observed lifetime of identified packets (whose dynamics are characterised herein) versus their intrinsic phase speed (Left) and mean orientation (Right). Each point represents a single packet. The error bars for intrinsic phase speed correspond to the velocity error, the calculation of which was performed with equation \ref{Vel_Error_eq}, while the error bars for orientation represent the standard deviation from the mean value between all orientation measurements for each specific packet.}
                \label{ABS_Icp_vs_Ori}
        \end{figure}
        We can see from Fig. \ref{ABS_Icp_vs_Ori} that most of the identified waves tend towards shorter lifetimes even for the case of IR2 waves. Even though few wave packets were observed over long periods (more than 10 hours), wave packets that live the longest tend to have slower intrinsic speeds. For shorter-lived packets, intrinsic phase speed is shown to be more variable. The right plot compares the observed packet lifetime with their respective mean orientations from parallel. Even though most packets have small orientations (mostly zonal downstream propagation) there is no apparent influence of orientation on lifetime, as shown by the data retrieved from IR2. The left plot possibly indicates that waves tend towards equilibrium with the zonal flow when they disperse \citep{Sutherland2010}.
        \subsection{Possible mechanisms for wave generation}
        \subsubsection{Surface forcing}
        The asymmetry of wave occurrence on Venus could suggest a forcing mechanism probably linked to either topography (non-stationary Lee waves) or any other localised features that are dependent on longitude or local time. In both bottom plots of Fig. \ref{NImages_Wave_Occurence_Panel}, there are two areas where wave occurrence is higher, one of which corresponds to the Aphrodite Terra, a large mountain range, while the other does not appear to be linked to any remarkable topographic feature. It is also relevant to point out that we also see a higher occurrence of waves at local times not long after dusk (terminator) and before dawn. However, results supporting a clear local time dependence remain inconclusive, as already found by \citet{Peralta2008}.\\
        Even though there is a concentration of packets in the region between Helen Planitia and Lavinia Planitia this can be attributed to an observation bias as this region features a higher number of observations, especially with VIRTIS images during the observed period. As such, there is no clear wave dependence with any geographical location on Venus, at least for small-scale waves in the lower cloud.\\
        The static stability profile of the lower cloud and below has been observed from entry probes (Vega 2 down to the surface and Pioneer Venus probes down to 12 km altitude) and is available in the VIRA model \citep{Seiff1985, Zasova2007}. It shows a low(down to zero)-static-stability region between roughly 20 and 30 km altitude, and the zero-static-stability convective layer in the low and middle clouds (roughly 50-55 km altitude). This convective layer and the stable region immediately below is also characterised from the radio-occultations from Magellan, Pioneer Venus, and Akatsuki, as recently discussed by \citet{Ando2020}. It is clear from these observations that the static-stability profile is dependent on local time and latitude.\\
        As atmospheric gravity waves cannot propagate in unstable regions \citep{Nappo2002, Sutherland2010}, the vertical movement of waves that would be forced on the surface would be compromised through this low-static-stability region. However, \citet{Lefevre2020} showed that stationary gravity waves generated by topographical features can indeed travel upward to the cloud deck through a type of tunnelling effect due to their large vertical wavelengths. These stationary waves are proposed to explain the presence of the large-scale bow-shaped stationary waves observed with Akatsuki \citep{Fukuhara2017}. The transmission factor for waves with similar wavelengths to those retrieved for this study reaches 20\% for the lowest unstable layer (mixed layer) and up to 45\% for the cloud convective layer considering their thickness \citep{Lefevre2020}. Therefore, it could be possible that waves generated near the surface could be part of those seen in the lower cloud region in our work. Furthermore, the horizontal wavelength of trapped lee waves on Venus with the mesoscale model described in \citet{Lefevre2020} is about 150 km, which is ten times greater than what is found on Earth \citep{Ralph1997} and consistent with the waves in this study. However, according to the simulations performed by \citet{Lefevre2020}, the vertical wavelengths of these waves should be at least three times greater than what we calculate from our estimation with radio occultation data. Additionally, such mountain-generated waves as those described in the observations performed in \citet{Kouyama2017} and the models of \citet{Lefevre2020} seem to be preferentially generated in the afternoon, which makes  observations of trapped lee waves generated by mountains on the nightside unlikely. Also, there are a significant number of waves whose location does not match any remarkable topography, and the mesoscale simulations imply that some of the estimated vertical wavelengths for the observed packet should not be allowed to propagate because of limitations from near-surface conditions. In addition, so far, stationary waves have not yet been reported on the nightside lower cloud \citep{Peralta2017b, Peralta2019}, preventing confirmation of this hypothesis, and the mesoscale simulations of orographic gravity waves might not be suitable for the non-stationary waves which are presented in this work.
        \subsubsection{Convection and Instabilities}
        It has been argued before that the most likely source of excitation for these waves is convection, in particular from the convectively unstable region in the lower and middle cloud \citep{Baker2000Part1, Baker2000Part2, Imamura2014}. Efforts to model convection-generated waves in the lower cloud \citep{McGouldrick2008} have shown that such waves could be observable in VEx/VIRTIS images. The most recent simulations done with an idealised Large-Eddy-Simulation model \citep{Lefevre2018} have shown that gravity waves were generated both above and below the convective layer. These latter authors also showed the strong influence of the vertical wind shear on the wavelengths and direction of propagation of the gravity waves. In this latter model, the strong convective activity induces gravity waves below the clouds over roughly 5 km, with or without wind shear. The presence of wind shear makes the wavefronts align perpendicularly to the wind direction, and increases the horizontal wavelength. This is interpreted by the authors as the consequence of an obstacle effect due to the interaction of the background wind with convective updrafts and downdrafts. This is consistent with the observed orientation measured in the present work, as well as with observations reported for the upper cloud (where the meridional wind is much stronger than below the clouds). Indeed our results are consistent with some predictions from models of convectively generated waves, namely their estimated vertical wavelength and spatial scales (morphological properties).\\
        Some of the characterised packets might also be generated through a Kelvin-Helmholtz instability mechanism. With the estimated values of the Brunt V\"{a}is\"{a}l\"{a} frequency from the static-stability profiles and the vertical profiles of the zonal wind from VIRA models presented in \citet{Peralta2014}, we calculated the Richardson number ($R_i$) for different latitudinal bins and heights as presented in \citet{SanchezLavegabook}, but without the contribution from the vertical shear of the meridional wind as we lack the spatial resolution to retrieve reliable data for the meridional wind in the lower cloud \citep{Hueso2012}. For an altitude range of 44-52 km, $R_i$ is mostly between 0.001 and 0.194 in our calculations for Venus, which suits the narrow region for the generation of Kelvin-Helmholtz instabilities (0 < $R_i$ < 0.25) \citep{SanchezLavegabook}. However, the critical value of $R_i$ for which instabilities occur is subject to debate because some authors suggest that the flow might be unstable for much higher values of $R_i$ \citep{Piccialli2010}, leading to some uncertainty over the conditions where we might expect these types of instabilities to form and whether or not these can be responsible for gravity waves such as those characterised in this paper.\\
        Some of these waves could also be generated by shear instabilities within the lower cloud. Some of these packets seemingly interact with their environment and such perturbations could also be part of a wave-generator mechanism.
        \begin{figure}[!h]
                \centering
                \includegraphics[width=1.0\columnwidth]{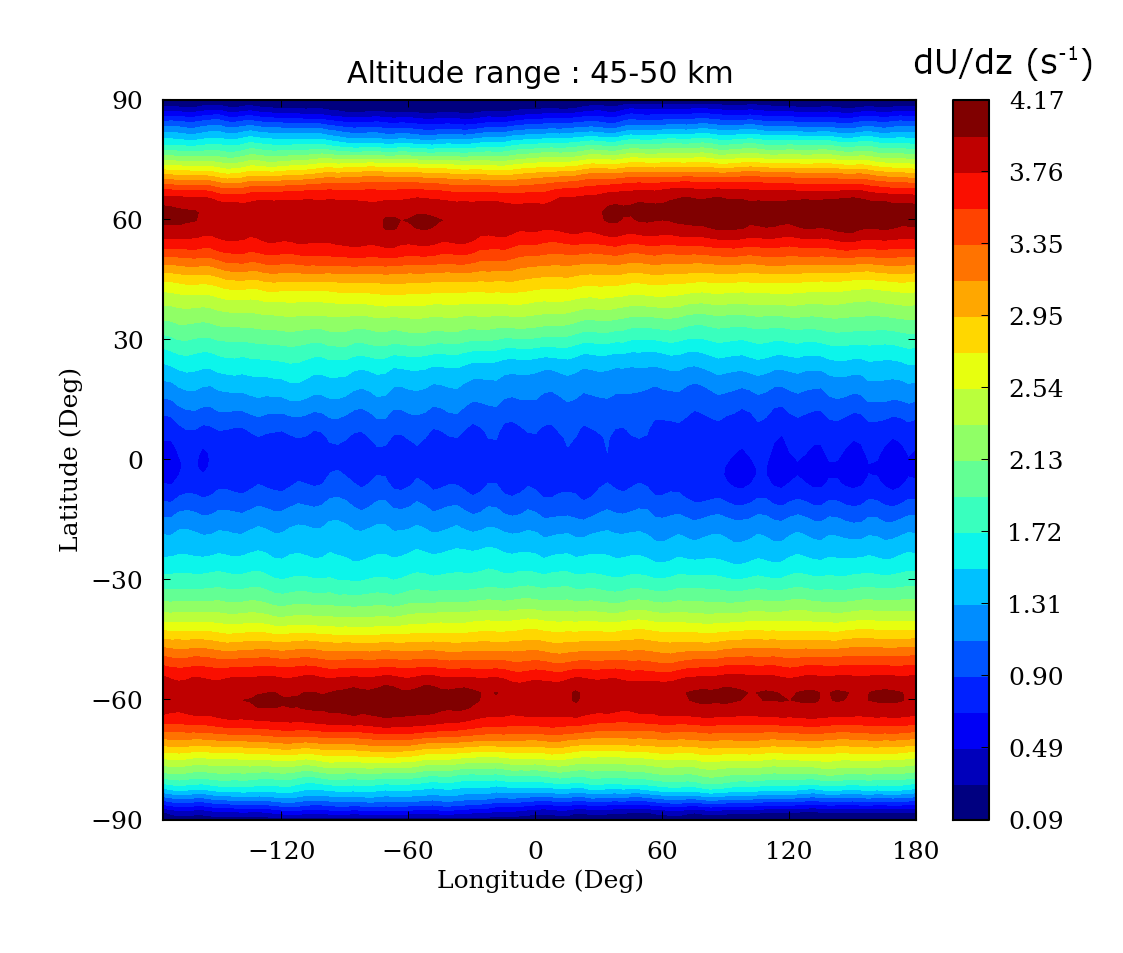}
                \caption{\footnotesize Latitude/longitude map of the vertical shear of the zonal wind at the altitude of the lower cloud layer. Vertical wind shear is in m.s$^{-1}$.Km$^{-1}$. An increase in the vertical shear is noticeable in the cold collar region on both hemispheres.\label{map_uwindshear}}
        \end{figure}
        From Fig. \ref{map_uwindshear} we notice there is an increase in the vertical shear of the zonal wind at latitudes close to 60$^{\circ}$. This not only coincides with the cold collar region but also with the highest concentration of packets observed in these observations. Even though this large number of packets has been addressed as an observational bias from VIRTIS, their location might be linked to this rise in wind shear and be generated by an instability provoked from it.\\[0.2cm]
        One last aspect regarding wave generation and propagation that is important to clarify is that, depending on the source, waves in this cloud region travel in vertically different directions, therefore either propagating towards the sky, but into the convective layer with almost zero static stability, or to deeper layers of the atmosphere of Venus \citep{Lefevre2018}.
        \subsection{Wave excitation and dispersion: Impact on circulation}
        Observation of apparent packet excitation and/or dissipation was possible for a small number of packets. These were registered simply as positive confirmation of the presence of a wave in a given location and, with the calculated value of its phase velocity, we can estimate its position on another image at an earlier or later date. In a very small number of cases, we were able to catch a glimpse of either small disturbances that would grow into a wave packet that was characterised, the result of breaking or dissipation of a wave, or simply the disappearance of a packet in the time interval between two sequential images of the same location. Some of these packets are seen to lose part of their structure, and in one particular case we see a wave packet interacting with another feature in the atmosphere of Venus and its structure being dissipated as it passes through. We analysed the influence of the wave packet dissipation on the background wind flow by taking wind tracers in the region where the wave would pass before or when the wave packet was active and after breaking.\\
        \begin{table*}
                \centering\footnotesize
                \caption{Observed appearance/disappearance of wave packets}
                \begin{tabular}{cccccccc}
                        \hline\hline
                        \noalign{\vspace*{0.2cm}}
                        Orbit & Date & Time & Lat & LT & $\hat{c_p^x}$ & $\delta$U & $\Delta$U\\
                        & (dd/mm/yyyy) & (UT-hour) & (deg) & (hour) & \multicolumn{3}{c}{(m/s)}\\
                        \noalign{\vspace*{0.2cm}}
                        \hline
                        \noalign{\vspace*{0.2cm}}
                        607 & 18/12/2007 & 17.48 - 22.48 & -44 & 2.7 & -1.575 & 11.041 & 14.241 \\[0.1cm]
                        r0025 & 04/09/2016 & 15.06 - 17.06 & 1 & 1.6 & 3.901 & 1.553 & 3.884 \\[0.1cm]
                        r0026 & 05/09/2016 & 03.56 - 05.56 & -24 & 0.8 & -1.741 & 3.439 & 14.844 \\[0.1cm]
                        r0026 & 05/09/2016 & 04.56 - 09.56 & -32 & 23.3 & -4.667 & 3.846 & 18.38 \\[0.1cm]
                        r0026 & 05/09/2016 & 04.56 - 10.56 & -40 & 23.1 & 2.131 & 3.509 & 15.519 \\[0.2cm]
                        \hline
                \end{tabular}
                \tablefoot{The orbit column follows the respective nomenclature norms for each spacecraft (the first entry corresponds to VEx/VIRTIS and the four remaining belong to Akatsuki/IR2); Time refers to the temporal window of observation of the wave packet; Lat, LT refer to the mean latitude and local time of the packet during propagation; $\hat{c_p^x}$ is the intrinsic phase velocity; $\delta$U is the wind measurement error and $\Delta$U is the wind speed drop between the wave packet appearance and disappearance.}
                \label{WaveDissipationTable}
        \end{table*}
        Even though the breaking of gravity waves dumps energy and momentum on their respective atmospheric layer, and as such we should expect an increase in the background wind flow velocity after dissipation, we verified that for all cases the wind flow was slowed after the wave disappeared and in four of the five cases we could see that for waves with greater intrinsic phase speed, this drop was larger (see Table \ref{WaveDissipationTable}).\\
        Unfortunately, opportunities to accompany wave propagation until breaking or dissipation were extremely rare for this data set because this has to be combined with the already limited available data of dynamically characterised waves as explained in Sect.\ref{Characterisation_DynamicalProp}. Possible solutions might include more continuous observations of the nightside of Venus such as those that could hypothetically be achieved by what is proposed in \citet{Kovalenko2020}, with micro-spacecraft inserted on Sun-Venus' Lagrange point orbits or further model studies on the transmission of waves between layers of the atmosphere similar to \citet{Lefevre2020}, concentrating on mesoscale waves in order to distinguish exactly where these waves dissipate.
        \subsection{Comparison with previous wave studies}
        Beyond the above-mentioned study of atmospheric waves in the lower cloud \citep{Peralta2008}, very few extensive analyses of this kind have been carried out. \citet{Peralta2019} characterised a large number of cloud morphologies observed on the nightside of Venus with Akatsuki/IR2 including wave packets. Even if they were not the focus of that particular study, waves identified in \citet{Peralta2019} served as valuable confirmation of wave packets identified for this work as well as validation of some characterisations of waves. Another study, focusing on atmospheric waves on the dayside upper clouds, was presented by \citet{Piccialli2014}. Their survey covers the northern hemisphere from 45$^{\circ}$ to polar latitudes at a latitude of approximately 66 km. Such waves are also interpreted as gravity waves, however the morphological properties of those wave packets have much smaller scales which can be related to the narrower field of view and higher spatial resolution of VEx/VMC images during pericentric observations. The dimensions of the upper cloud waves in this latter work were reported to be about one order of magnitude smaller than what we find for wave packets in the lower cloud. The orientation of wave packets has a broader distribution than our findings and wave packets are concentrated above the mountain range Ishtar Terra. Wave properties and the distribution of wave activity suggests that Kelvin-Helmholtz instabilities or surface forcing play important roles in generating the waves found by \citet{Piccialli2014} and \citep{Peralta2019}.\\
        The contrast between dayside upper cloud and nightside lower cloud waves is readily apparent by their morphological properties and distribution, even if the study by \citet{Piccialli2014} was confined to the region northward of 45$^{\circ}$ due to an observational bias from the Venus Monitoring Camera (VEx/VMC) \citep{Markiewicz2007} used in their survey. Also, because of the spacecraft orbit, these latter authors were unable to retrieve dynamical properties from waves in the upper cloud \citep{Moissl2009}.\\
        This divergence in properties could also indicate a different forcing mechanism for gravity waves at different altitudes, or it could be that the conditions in which we find both types of wave packets (dayside upper cloud and nightside lower cloud) constrain the observable morphological properties. Future analyses of wave packets in the upper clouds in ultraviolet images from Akatsuki's Ultraviolet Imager (UVI) could help further confirm these hypotheses and possibly establish a connection between gravity waves on both cloud layers.
        
        \section{Conclusions}
        
        We present a systematic study of mesoscale atmospheric waves on the nightside of Venus with data from the VIRTIS-M-IR channel and IR2 from Venus Express space mission and Akatsuki respectively. The wave packets apparent in the images were morphologically and dynamically characterised, being interpreted as atmospheric gravity waves propagating on the lower clouds of Venus. This paper serves as a follow-up work from what is discussed in \citet{Peralta2008}, but focuses on nightside waves only and uses two different data sets to expand the search for waves to other locations besides what was possible with Venus Express, and is the largest observational study of waves on the nightside of Venus to date. We also extend the preliminary characterisation in Akatsuki images presented by \citet{Peralta2019}.\\[0.2cm]
        Atmospheric waves were mostly detected on the southern hemisphere of Venus, which is probably due to the observation bias of VIRTIS data, as there is no evidence from previous exploits or in our data to suggest that there would be a significant asymmetry on wave generation between both hemispheres. However, this can only be verified with further observations of the northern hemisphere of Venus at the lower cloud.\\[0.2cm]
        Waves detected with VIRTIS and IR2 show similar characteristics regarding their morphological and dynamical properties, though IR2 waves show more variety even with less distinct packets. This could be attributed to waves being detected on a wider range of latitudes and local times, especially closer to equatorial latitudes. Further evaluation would be required to examine the interaction between the different flow regimes on the lower cloud of  Venus and the properties of waves. However, we speculate that this could be related to different forcing mechanisms at hand.\\[0.2cm] 
        On another note, we verify that the general background wind between both data sets increases by roughly 10 m/s. As wind retrievals from IR2 are more concentrated at lower latitudes (equatorward) where the transient zonal wind jet \citep{Horinouchi2017} is more prevalent and the atmosphere is thicker in general, higher zonal wind values would be expected which we can see in Fig. \ref{ZonalWindProfile_Waves}. However, the dynamics of the lower cloud of Venus have been seen to change \citep{Peralta2018}, as have those of the upper cloud \citep{Hueso2015}.
        Wave phase velocity and trajectory suggest that these waves are `guided' by the background zonal wind flow given their low intrinsic speeds and orientations. We observe a decrease in the local wind speed after waves dissipate but the short number of data points where this was verifiable does not allow for a more robust interpretation. However, it is apparent that gravity waves are restricted by either their forcing mechanisms or the background dynamics of Venus to low intrinsic phase speed and orientations.\\[0.2cm]
        We argue that convection is the main driving force of these waves but it is still not possible to rule out other sources of wave generation such as topography, shear, or Kelvin-Helmholtz instabilities. We hope that the data presented here, providing an update of direct measurements of gravity waves, can be used by recent and future models to better predict the influence of atmospheric gravity waves on the general circulation of Venus' atmosphere.
        
        \begin{acknowledgements}

        This research is supported by the University of Lisbon through the BD2017 program based on the regulation of investigation grants of the University of Lisbon, approved by law 89/2014, the Faculty of Sciences of the University of Lisbon and the Portuguese Foundation for Science and Technology FCT through the project P TUGA PTDC/FIS-AST/29942/2017. We also acknowledge the support of the European Space Agency and the associated funding bodies Centre National d’Etudes Spatiales (France) and Agenzia Spaziale Italiana (Italy) as well as the full team behind the VIRTIS instrument, Venus Express space mission and the PSA archives. Additionally, we acknowledge the support and work of the entire Akatsuki team. The first author also acknowledges the full support of Japan Aerospace Exploration Agency (JAXA) for enabling a short internship in their facilities which greatly contributed to this work. J.P. acknowledges JAXA's International Top Young Fellowship (ITYF). M.L. acknowledges funding from the European Research Council (ERC) under the European Union's Horizon 2020 research and innovation program (grant agreement No. 740963/EXOCONDENSE)
        \end{acknowledgements}

\bibliographystyle{aa}
\bibliography{BibliographyGeneral}
\end{document}